\documentclass[12pt,onecolumn]{IEEEtran} 

\usepackage{graphicx}  
\usepackage{psfrag}    
\usepackage{subfigure} 
\usepackage{amsmath}   
\usepackage{amssymb}   
\usepackage{eucal}     

\newtheorem{proposition}{Proposition}
\newtheorem{definition}{Definition}
\newtheorem{corollary}{Corollary}

\newtheorem{theorem}{Theorem}
\newcommand{\mbI}{{\mathbb{I}}}
\newcommand{\mbH}{{\mathbb{H}}}

\DeclareMathAlphabet{\eurm}{U}{eur}{m}{n}
\DeclareMathAlphabet{\mathbsf}{OT1}{cmss}{bx}{n}
\DeclareMathAlphabet{\mathssf}{OT1}{cmss}{m}{sl}
\DeclareMathAlphabet{\mathcsf}{OT1}{cmss}{sbc}{n}

\newcommand{\samplevalue}[1]{\eurm{\lowercase{#1}}}
\newcommand{\randomvalue}[1]{\eurm{\uppercase{#1}}}

\DeclareSymbolFont{bsfletters}{OT1}{cmss}{bx}{n}  
\DeclareSymbolFont{ssfletters}{OT1}{cmss}{m}{n}
\DeclareMathSymbol{\bsfGamma}{0}{bsfletters}{'000}
\DeclareMathSymbol{\ssfGamma}{0}{ssfletters}{'000}
\DeclareMathSymbol{\bsfDelta}{0}{bsfletters}{'001}
\DeclareMathSymbol{\ssfDelta}{0}{ssfletters}{'001}
\DeclareMathSymbol{\bsfTheta}{0}{bsfletters}{'002}
\DeclareMathSymbol{\ssfTheta}{0}{ssfletters}{'002}
\DeclareMathSymbol{\bsfLambda}{0}{bsfletters}{'003}
\DeclareMathSymbol{\ssfLambda}{0}{ssfletters}{'003}
\DeclareMathSymbol{\bsfXi}{0}{bsfletters}{'004}
\DeclareMathSymbol{\ssfXi}{0}{ssfletters}{'004}
\DeclareMathSymbol{\bsfPi}{0}{bsfletters}{'005}
\DeclareMathSymbol{\ssfPi}{0}{ssfletters}{'005}
\DeclareMathSymbol{\bsfSigma}{0}{bsfletters}{'006}
\DeclareMathSymbol{\ssfSigma}{0}{ssfletters}{'006}
\DeclareMathSymbol{\bsfUpsilon}{0}{bsfletters}{'007}
\DeclareMathSymbol{\ssfUpsilon}{0}{ssfletters}{'007}
\DeclareMathSymbol{\bsfPhi}{0}{bsfletters}{'010}
\DeclareMathSymbol{\ssfPhi}{0}{ssfletters}{'010}
\DeclareMathSymbol{\bsfPsi}{0}{bsfletters}{'011}
\DeclareMathSymbol{\ssfPsi}{0}{ssfletters}{'011}
\DeclareMathSymbol{\bsfOmega}{0}{bsfletters}{'012}
\DeclareMathSymbol{\ssfOmega}{0}{ssfletters}{'012}

\newcommand{\rvq}{{\randomvalue{q}}}	
\newcommand{\svq}{{\samplevalue{q}}}

\newcommand{\rvs}{{\randomvalue{s}}}	
\newcommand{\svs}{{\samplevalue{s}}}

\newcommand{\rvu}{{\randomvalue{u}}}	
\newcommand{\svu}{{\samplevalue{u}}}	

\newcommand{\rvw}{{\randomvalue{w}}}	
\newcommand{\svw}{{\samplevalue{w}}}

\newcommand{\rvx}{{\randomvalue{x}}}	
\newcommand{\svx}{{\samplevalue{x}}}

\newcommand{\rvy}{{\randomvalue{y}}}	
\newcommand{\svy}{{\samplevalue{y}}}

\newcommand{\rvz}{{\randomvalue{z}}}	

\newcommand{\calA}{{\mathcal{A}}}

\newcommand{\calC}{{\mathcal{C}}}

\newcommand{\calN}{{\mathcal{N}}}

\newcommand{\calP}{{\mathcal{P}}}
\newcommand{\calQ}{{\mathcal{Q}}}
\newcommand{\calR}{{\mathcal{R}}}

\newcommand{\calS}{{\mathcal{S}}}
\newcommand{\calU}{{\mathcal{U}}}

\newcommand{\calX}{{\mathcal{X}}}
\newcommand{\calY}{{\mathcal{Y}}}
\newcommand{\calW}{{\mathcal{W}}}

\begin{document}

\title{Multiaccess Channels with State Known 
to Some Encoders and Independent Messages}


\author
{{Shiva~Prasad Kotagiri and  J. Nicholas Laneman.}
\thanks{Part of this work was published in Allerton 
Conference on Communications and Control, Monticello, IL, USA, October 2004}
\thanks{This work has been supported in part by NSF grants CCF05-46618 and  
CNS06-26595, the Indiana 21st Century Fund, and a Graduate Fellowship  
from the Center for Applied Mathematics at the University of Notre  
Dame}
 \thanks{ShivaPrasad Kotagiri and J. Nicholas Laneman 
 are with Department of Electrical Engineering,
 University of Notre Dame, Notre Dame, IN 46556, 
 Email: \texttt{skotagir@gmail.com, jnl@nd.edu}}}

\maketitle

\begin{abstract}
We consider a state-dependent multiaccess channel (MAC) with state 
\textit{non-causally} known  to some encoders.
For simplicity of exposition, we focus on a two-encoder model in which
\textit{one} of the encoders has  non-causal access to the channel state. 
The results can in
principle be extended to any number of encoders with a subset of them
being informed. We derive an inner bound for the capacity region
in the general discrete memoryless case and specialize to a binary
noiseless case. In binary noiseless case, we compare the inner bounds with trivial outer bounds obtained by 
providing the noiseless channel state to the decoder. In the case of maximum entropy channel state, we obtain the
capacity region for binary noiseless MAC with one informed encoder by deriving a non-trivial 
outer bound for this case.

For a Gaussian state-dependent MAC with one encoder
being informed of the channel state, we present an inner bound by applying a slightly generalized dirty paper coding (GDPC) at the informed encoder that allows 
for partial state cancellation,  and a trivial outer bound by providing channel state to the decoder also.  In particular, if the channel input is negatively 
correlated with the channel state in the random coding distribution, 
then GDPC can be interpreted as partial state cancellation followed by standard dirty paper coding. 
The uninformed encoders benefit 
from the state cancellation in terms of achievable rates, however, appears that GDPC cannot completely
eliminate the effect of the channel state on the achievable rate region, in contrast to the case of
all encoders being informed. In the case of infinite state variance, we analyze how the uninformed encoder benefits from the informed encoder's actions using the inner bound and also provide a non-trivial outer bound for this case which is better than the trivial outer bound. 
\end{abstract}


%
\IEEEpeerreviewmaketitle

\section{Introduction} \label{sec:introduction}
\begin{figure}[h]
\begin{center}
\psfrag{IE}[cc]{Informed Encoder}
\psfrag{UE}[cc]{Uninformed Encoder}
\psfrag{MAC}[cc]{Multiaccess Channel}
\psfrag{SG}[cc]{State Generator}
\psfrag{DE}[cc]{Decoder}
\psfrag{X1}[cc]{$\rvx_{1}^{n}$}
\psfrag{W1}[cc]{$\rvw_1$}
\psfrag{W2}[cc]{$\rvw_2$}
\psfrag{X2}[cc]{$\rvx_{2}^{n}$}
\psfrag{S}[cc]{$\rvs^{n}$}
\psfrag{Y}[cc]{$\rvy^{n}$}
\psfrag{D}[cc]{$(\hat{\rvw}_1,\hat{\rvw}_2)$}
\psfrag{P}[cc]{$p(\svy|\svx_1,\svx_2,\svs)$}
\resizebox{0.8\columnwidth}{!}{
\includegraphics{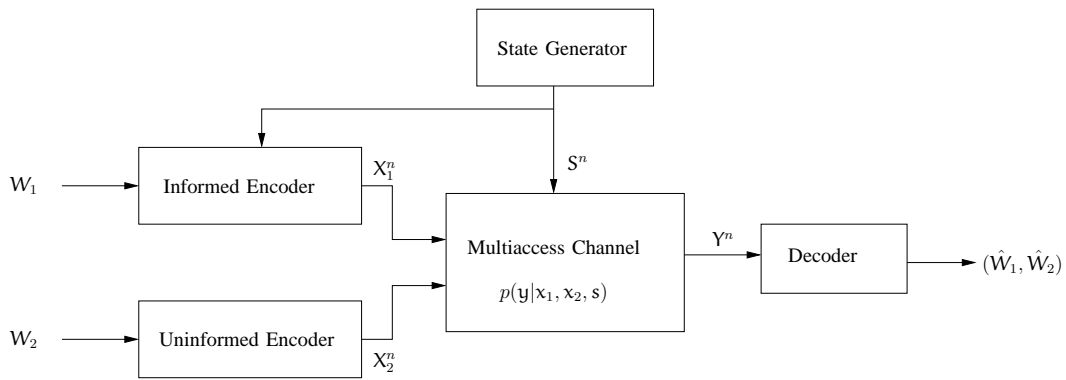}}
\end{center}
\caption{State-dependent multiaccess channel with channel state 
non-causally known to one encoder.} \label{fig:mac}
\end{figure}

We consider a state-dependent multiaccess channel (MAC) with noiseless channel state
non-causally known to only some, but not all, encoders.  The
simplest example of a communication system under investigation is
shown in Figure~\ref{fig:mac}, in which two encoders communicate to a
single decoder through a state-dependent MAC $p(\svy|\svx_1, \svx_2,\svs)$ controlled 
by the channel state $\rvs$. We assume that one of the encoders has non-causal 
access to the noiseless channel state. The results can in principle be extended to any number of
encoders with a subset of them being informed of the noiseless channel state.
The informed encoder and the uninformed encoder want to send messages $\rvw_1$ and $\rvw_2$, respectively,
to the decoder in $n$ channel uses.
The informed encoder, provided with both $\rvw_1$ and the channel state
$\rvs^n$, generates the codeword $\rvx_1^n$. The uninformed encoder,
provided only with $\rvw_2$, generates the codeword
$\rvx_2^n$. The decoder, upon receiving the channel output $\rvy^n$,
estimates both  messages $\rvw_1$ and $\rvw_2$ from $\rvy^n$.
In this paper, our goal is to study the capacity region of this model.

\subsection{Motivation}
State-dependent channel models with state available at the encoder can be used to model IE 
\cite{bchen00,chen01:it,moulin03:it,acohen02:it}.
Information embedding (IE) is a recent area of digital media research
with many applications, including: passive and active copyright
protection (digital watermarking); embedding important control,
descriptive, or reference information into a given signal;
 and covert communications \cite{swanson98:picc}.
IE enables encoding a message
into a host signal (digital image, audio, video) such that it is
perceptually and statistically undetectable.  Given the various
applications and advantages of IE, it is important
to study fundamental performance limits of these schemes.
The information theory community has been studying performance limits
of such models in which random parameters capture  fading in a wireless environment,
 interference from other users \cite{caire03:it}, or the host sequence in IE and date
hiding applications \cite{bchen00,chen01:it,moulin03:it,acohen02:it,kalker02:picdsp}.

The state-dependent models with channel
state available at the encoders can also be used to model communication systems with 
cognitive radios. Because of growing demand for bandwidth in wireless systems, 
some secondary users with \textit{cognitive}
 capabilities are introduced into an existing primary communication 
system to use the frequency spectrum more efficiently \cite{mitola00:phd}. These cognitive devices  
are supposed to be capable of obtaining knowledge about the primary communication that takes place in the 
channel and adapt their coding schemes to remove the effect of interference caused by the primary 
communication systems to increase  spectral efficiency.  
The state in such models can be viewed as the signal of the primary communication that 
takes place in the same channel, and the informed encoders can be viewed as cognitive users. 
The model considered in the paper can be viewed as a secondary multiaccess communication system with 
some cognitive and non-cognitive users introduced into the existing primary communication system. 
The cognitive users  are capable of non-causally obtaining the channel state or the signal of the
 primary communication system. 
In this paper, we are interested in studying the achievable rates of the secondary multiaccess communication 
system with some cognitive users. Joint design of the primary and the secondary networks is studied 
in \cite{devroye06:it,jovicic06:it}.

\subsection{Background}
The study of state-dependent models or channels with random parameters, 
primarily for single-user channels, is initiated with Shannon himself. Shannon studies the single-user
discrete memoryless (DM) channels $p(\svy|\svx,\svs)$ with causal channel state at the encoder
\cite{shannon58:ibmj}. Here, $\rvx$, $\rvy$, and $\rvs$ are the channel input,
output, and state respectively. Salehi studies the capacity of
these models when different noisy observations of the channel state are causally
known to the encoder and the decoder \cite{salehi92:piee}. Caire and
Shamai extend the results of \cite{salehi92:piee} to channels with
memory \cite{caire99:it}.

Single-user DM state-dependent channels with memoryless state non-causally
known to the encoder are studied in \cite{kusnetsov74:ppi,cheegard83:it}
in the context of computer memories with defects.
Gel'fand-Pinsker derive the capacity of these models, which is given by  \cite{gelfand80:pcit}
\begin{equation}
C= \max_{p(\svu|\svs),~ \rvx=f(\rvu,\rvs)} [\mathbb{I}(\rvu;\rvy)-\mathbb{I}(\rvu;\rvs)]
\end{equation}
where $\rvu$ is an auxiliary random variable, and $\rvx$ is a 
deterministic function of $(\rvu,\rvs)$. Single- user DM channels 
with two state components, one component non-causally known to the encoder
and another component known to the decoder, are studied in~\cite{cover02:it}.

Costa studies the memoryless additive white
Gaussian state-dependent channel of the form $\rvy^n =\rvx^n+\rvs^n+\rvz^n$,
where $\rvx^n$ is the channel input with power
constraint $\frac{1}{n}\sum_{i=1}^n \rvx_{i}^2 \leq P$,
$\rvs^n$ is the memoryless state vector whose elements are
non-causally known to the encoder and are
zero-mean Gaussian random variables with
variance $Q$, and $\rvz^n$ is
 the memoryless additive noise vector whose elements are zero mean
 Gaussian random variables
 with variance $N$ and are independent of the channel input and  the state.
 The capacity of this model is given by \cite{mcosta83:it}:
\begin{equation}
C = \frac{1}{2}\log \left ( 1+ \frac{P}{N} \right ). \label{eqn:costa_capacity}
\end{equation}
In terms of the capacity, the result (\ref{eqn:costa_capacity}) indicates that non-causal state at the encoder is equivalent to state at the decoder or no state in the channel.
The so-called \textit{dirty paper coding} (DPC)
scheme used to achieve the capacity (\ref{eqn:costa_capacity})
suggests that allocating power for explicit state cancellation is not optimal, i.e.,
the channel input $\rvx$ is uncorrelated with the channel state $\rvs$ in 
the random coding distribution
\cite{mcosta83:it}.

For state-dependent models with non-causal state at the encoder, although much
is known about the single user case, the theory is less well developed
for multi-user cases.  Several groups of researchers
\cite{gelfand83:pisit, kim04:pisit} study the memoryless additive
Gaussian state-dependent MAC of the form $\rvy^n =\rvx_1^n+\rvx_2^n+\rvs^n+\rvz^n$, where:
$\rvx_1^n$ and $\rvx_2^n$ are the channel inputs with average power
constraints $\frac{1}{n}\sum_{i=1}^n \rvx_{1,i}^2 \leq P_1$ and 
$\frac{1}{n}\sum_{i=1}^n \rvx_{2,i}^2 \leq P_2$,
respectively, $\rvs^n$ is the
memoryless channel state vector whose elements are non-causally known  at
 \textit{both} the encoders and are zero-mean Gaussian random variables with
variance $Q$, and $\rvz^n$
is the memoryless additive noise vector whose elements are zero-mean
Gaussian random variables with variance $N$ and are independent of the
channel inputs and the channel state. The capacity region of this model is the
set of rate pairs $(R_1,R_2)$ satisfying
{\allowdisplaybreaks
\begin{subequations}\label{eqn:gmac_capacity}
\begin{align}
R_1\leq & \frac{1}{2}\log \left (1+ \frac{P_1}{N}\right )   \\
R_2 \leq & \frac{1}{2}\log\left (1+ \frac{P_2}{N}\right )  \\
R_1+R_2 \leq & \frac{1}{2}\log\left (1+ \frac{P_1+P_2}{N}\right ).
\end{align}
\end{subequations}}
As in the single-user Gaussian model, the capacity region (\ref{eqn:gmac_capacity}) 
indicates that the channel state has no
effect on the capacity region if it is non-causally known to both the
encoders.  Similar to the single-user additive
Gaussian models with channel state, DPC at both the encoders achieves
(\ref{eqn:gmac_capacity}) and explicit state cancellation is not optimal in terms
of the capacity region. 
It is interesting to study the capacity region for the Gaussian MAC 
with non-causal channel state at one encoder because DPC cannot be applied at 
the uninformed encoder.

For the DM case, the state-dependent MAC with state at 
one encoder is considered in \cite{baruch06:pallerton, baruch07:pisit, kotagiri07:pisit} 
when the informed encoder knows the
message of the uninformed encoder. For the Gaussian case in the same scenario, the capacity region 
is obtained in \cite{baruch07:pisit,baruch07:it} by deriving non-trivial outer bounds. It is shown that the 
generalized dirty paper coding (GDPC) achieves the capacity region. 
The model considered in this paper from the view of lattice coding is also considered in 
\cite{philosof07:pisit}. Cemal and Steinberg study the state-dependent MAC in which
rate-constrained state is known to the encoders and full state is known to
the decoder \cite{ycemal05:it}.  State-dependent Broadcast channels with state available at
the encoder have also been studied in the DM case~\cite{steinberg05:pisit,ysteinberg05:it} 
and the Gaussian case \cite{khisti04:pisit}.

\subsection{Main Contributions and Organization of the paper}
We derive an inner bound for the model shown in
Figure~\ref{fig:mac} for the DM case and then specialize to a binary
noiseless case. General outer bounds for these models 
have been obtained in \cite{kotagiri08:phd}, however, at present, these bounds
do not coincide with our inner bounds and are not computable 
due to lack of bounds on the cardinalities of the auxiliary random variables. 
In binary noiseless case, the informed encoder uses a slightly
generalized binary DPC, in which the random coding distribution has channel input random variable correlated to 
the channel state. If the binary channel state is
a Bernoulli$(q)$ random variable with $q <0.5$, we compare the inner bound with a
trivial outer bound obtained by providing the channel state to the decoder, and 
the bounds do not meet. 
If $q=0.5$, we obtain the capacity region by deriving a non-trivial outer bound.

We also derive an inner bound for an additive
white Gaussian state-dependent MAC similar to \cite{gelfand83:pisit, kim04:pisit},
but in the asymmetric case in which one of the encoders has 
non-causal access to the state. 
For the inner bound, the informed encoder uses
a generalized dirty paper coding (GDPC) scheme in which the random coding distribution
exhibits arbitrary correlation between the channel input from the informed encoder and the channel state.
The inner bound using GDPC is compared with a trivial outer bound obtained 
by providing channel state to the decoder. If the channel input from the informed encoder is negatively 
correlated with the channel state, then GDPC can be interpreted as partial state cancellation followed by standard
dirty paper coding.  We observe that, in terms
of achievable rate region, the informed encoder with GDPC can assist the
uninformed encoders. However, in contrast to the case of channel state available
at all the encoders \cite{gelfand83:pisit, kim04:pisit}, it appears
that  GDPC cannot completely eliminate the effect of the channel state
on the capacity region for the Gaussian case.

We also study the Gaussian case if the channel state  has asymptotically large
variance $Q$, i.e., $Q \rightarrow \infty $. Interestingly, the uninformed
encoders can benefit from the informed encoder's actions.
 In contrast to the case of $Q < \infty$
in which the informed encoder uses GDPC, we
show that the standard DPC is sufficient to help
the uninformed encoder as $Q \rightarrow \infty $.
 In this latter case, explicit state cancellation  is not useful because it
 is impossible to explicitly cancel the channel state using the finite power of
 the informed encoder.

We organize the rest of the paper as follows. In Section~\ref{sec:notations},
 we define some notation and the capacity region.
In Section~\ref{sec:innerbound_dmcase}, we study a general inner bound for
the capacity region of the model in Figure~\ref{fig:mac} for a DM MAC and
also specialize to a  binary noiseless case. In this section, we also derive the capacity region of the binary noiseless MAC with maximum entropy channel state.
In Section~\ref{sec:gaussiancase}, we study inner and outer bounds on the capacity region of the model 
in Figure~\ref{fig:mac} for a memoryless Gaussian state-dependent MAC and also study the inner and outer bounds for the capacity region of this model in the case of large channel state variance. Section~\ref{sec:conclusions} concludes the paper.

\section{Notations and Definitions}\label{sec:notations}
Throughout the paper, the notation $\svx$ is used to denote the realization of
the random variable $\rvx \sim p(\svx)$. The notation
$\rvx_1^n$ represents the sequence $\rvx_{1,1},\rvx_{1,2},\ldots,\rvx_{1,n},$
and the notation $\rvx_{1,i}^n$ represents the sequence
$\rvx_{1,i},\rvx_{1,i+1},\ldots,\rvx_{1,n}$. Calligraphic letters are used to
denote the random variable's alphabet, e.g., $\rvx \in \calX$. The notation
$\mathrm{cl}\{\calA\}$ and $\mathrm{co}\{\calA\}$ denote the closure operation
and convex hull operation on set $\calA$, respectively.

We consider a  memoryless state-dependent MAC, denoted  $p(\svy|\svx_{1},\svx_{2},\svs)$, 
whose output $\rvy \in \calY$ is controlled by the channel input pair $(\rvx_1,\rvx_2) \in (\calX_1,\calX_2)$ and
along with the channel state $\rvs \in \calS$. These alphabets are discrete sets and the set of real numbers for the
discrete case and the Gaussian case, respectively. We assume that $\rvs_{i}$ at any time instant $i$
is identically independently drawn (i.i.d.) according to a probability law $p(\svs)$.
As shown in Figure~\ref{fig:mac}, the state-dependent MAC is embedded
in some environment in which channel state  is non-causally known  to one encoder.

The informed encoder, provided with the non-causal channel state, wants to send message $\rvw_1$ 
to the decoder and the uninformed encoder wants to send $\rvw_2$ to the decoder. 
The message sources at the informed encoder and the uninformed encoder
produce random integers
\[\rvw_{1}\in \{1,2,\ldots, M_{1}\} \quad  \text{and} \quad  \rvw_{2}\in
\{1,2,\ldots,M_{2}\}, \] respectively,
at the beginning of each block of $n$ channel uses.
 We assume that the messages are independent and the probability of each pair of messages
 $(\rvw_1=\svw_{1},\rvw_2=\svw_{2})$ is given by $\frac{1}{M_{1}M_{2}}$.

\begin{definition}
A $(\lceil 2^{nR_1}\rceil,\lceil2^{nR_2}\rceil,n)$  code consists of encoding functions
\[f_1^n : \mathcal{S}^n \times \calW_1 \rightarrow \mathcal{X}_1^n ~\mathrm{and}~
f_2^n : \calW_2 \rightarrow \mathcal{X}_2^n \]
 at the informed encoder and  the uninformed encoder, respectively,
 and a decoding function
\[g^n: \mathcal{Y}^n \rightarrow (\calW_1 \times \mathcal{W}_2),\]
where $\calW_i =\{1,2,\ldots,\lceil 2^{nR_i} \rceil\}$ for $i=1,2.$
\end{definition}

From a $(\lceil 2^{nR_1}\rceil,\lceil2^{nR_2}\rceil,n)$ code, the sequences
$\rvx_1^n$ and $\rvx_2^n$ from the informed encoder and
the uninformed encoder, respectively,
 are transmitted without feedback across a state-dependent MAC $p(\svy|\svx_1,\svx_2,\svs)$ modeled as a discrete
memoryless conditional probability distribution, so that
\begin{equation}
\mathrm{Pr}(\rvy^n=\svy^n|\svs^n,\svx_1^n,\svx_2^n)= \prod_{j=1}^{n}
p(\svy_j|\svx_{1,j},\svx_{2,j},\svs_{j}). \label{eqn:channellaw}
\end{equation}
The decoder, upon receiving the channel output $\rvy^n$, attempts to reconstruct the messages.
The average probability of error is defined as
$ P_e^n = \mathrm{Pr}[g(\rvy^n)\neq (\rvw_1,\rvw_2)].$

\begin{definition}
A rate pair $(R_1,R_2)$ is said to be achievable
if there exists
a sequence of $(\lceil 2^{nR_1}\rceil,\lceil2^{nR_2}\rceil,n)$  codes
$(f_1^n,f_2^n,g^n)$ with $\lim_{n \rightarrow \infty} P_e^n=0.$
\end{definition}

\begin{definition}
The capacity region $\calC$  is
the closure the set of achievable
rate pairs $(R_1,R_2)$.
\end{definition}

\begin{definition}
For given $p(\svs)$ and $p(\svy|\svx_1,\svx_2,\svs)$, let $\calP^i$ be
the collection of
random variables $(\rvq,\rvs,\rvu_1,\rvx_1,\rvx_2,\rvy)$
with probability laws
\begin{align}
p(\svq,\svs,\svu_1,\svx_1,\svx_2,\svy)  = & p(\svq)p(\svs)
p(\svu_1|\svs,\svq)p(\svx_1|\svu_1,\svs,\svq)p(\svx_2|\svq) \nonumber \\
 & \times p(\svy|\svx_1,\svx_2,\svq), \nonumber
\end{align}
where $\rvq$ and $\rvu_1$ are auxiliary random variables.
\end{definition}

\section{Discrete Memoryless Case} \label{sec:innerbound_dmcase}
In this section, we derive an inner bound for the capacity region of the model shown in Figure~\ref{fig:mac} for a general DM MAC and then specialize to a  binary noiseless MAC.
In this section,  we consider $\calX_1$, $\calX_2$, $\calS$, and
$\calY$ to all be discrete and finite alphabets, and all probability distributions are
to be interpreted as probability mass functions.

\subsection{Inner Bound for the Capacity Region}
The following theorem provides an inner bound for the
DM case.

\begin{theorem} \label{thm:dm_innerbound}
Let $\calR^i$ be the closure of all rate pairs $(R_1,R_2)$
satisfying
\begin{subequations} \label{eqn:dm_innerbound}
\begin{align}
R_{1} < & \mbI(\rvu_1; \rvy | \rvx_2,\rvq)-\mbI(\rvu_1;\rvs|\rvq) \\
R_2 < & \mbI(\rvx_2; \rvy| \rvu_1,\rvq) \\
R_1 + R_2 < & \mbI(\rvu_1, \rvx_2; \rvy|\rvq)-\mbI(\rvu_1;\rvs|\rvq)
\end{align}
\end{subequations}
for some random vector $(\rvq,\rvs,\rvu_1,\rvx_1,\rvx_2,\rvy)
\in \calP^i$, where $\rvq \in \calQ$ and $\rvu_1 \in \calU_1$ are auxiliary random variables with
$|\calQ|\leq 4$ and $|\calU_1| \leq |\calX_1||\calX_2||\calS|+4$, respectively.
Then the capacity region $\calC$ of the DM MAC with one informed encoder satisfies
$\calR^i \subseteq \calC$.
\end{theorem}

\noindent \textbf{Proof:} The above inner bound can be proved by essentially combining random channel coding 
for the DM MAC \cite{cover91:book} and random channel coding with non-causal state at the encoders 
\cite{gelfand80:pcit}. For completeness, a proof using joint decoding is given in Appendix~\ref{proof_innerbound}.

\noindent \textit{Remarks:}
\begin{itemize}
\item The inner bound of Theorem~\ref{thm:dm_innerbound} can be obtained by
applying Gel'fand-Pinsker
coding \cite{gelfand80:pcit} at the informed encoder.
At the uninformed encoder, the codebook is generated in
the same way as for a regular DM MAC \cite{cover91:book}.
\item The region $\calR^i$ in Theorem~\ref{thm:dm_innerbound} is convex due
to the auxiliary time-sharing random variable $\rvq$.
\item The inner bound $\calR^i$ of Theorem~\ref{thm:dm_innerbound} can also be obtained 
by time-sharing between two successive decoding schemes, i.e.,
decoding one encoder's message first and using the
decoded codeword and the channel output to decode the other encoder's message.
On one hand consider first
decoding the message of the informed encoder. Following \cite{gelfand80:pcit},
if $R_1 < \mbI(\rvu_1 ; \rvy)-\mbI(\rvu_1 ;\rvs)$,
 we can decode the codeword $\rvu_1^n$ of the informed encoder with arbitrarily
 low probability of error.
 Now, we use $\rvu_1^n$ along with $\rvy^n$ to decode $\rvx_2^n$. Under these
 conditions, if $R_2 < \mbI(\rvx_2;\rvy|\rvu_1)$, then we can decode the message
 of the uninformed
 encoder with arbitrarily low probability of error.
 On the other hand, if we change the decoding order of the two messages, the constraints are
 $R_2 < \mbI(\rvx_2;\rvy)$ and $R_1 < \mbI(\rvu_1;\rvy|\rvx_2)-\mbI(\rvu_1;\rvs)$.
 By time-sharing between these two successive decoding schemes and taking the convex closure, 
 we can obtain the inner bound 
 $\calR^i$ of Theorem~\ref{thm:dm_innerbound}.
\end{itemize}

\subsection{Binary Noiseless Example} \label{sec:binarycase}
In this section, we specialize Theorem~\ref{thm:dm_innerbound} to a binary noiseless
state-dependent MAC of the form
$\rvy^n=\rvx_{1}^n \oplus \rvx_{2}^n \oplus \rvs^n,$ where: 
$\rvx_{1}^n$ and $\rvx_{2}^n$ are channel inputs with the number of binary ones in
$\rvx_{1}^n$ and $\rvx_{2}^n$ less than or equal to $np_1$, $0 \leq p_1 \leq 1$, and $np_2$, $0 \leq p_2 \leq 1$, respectively; $\rvs^n$ is the memoryless state vector whose elements are
non-causally known to one encoder  and are i.i.d.  
Bernoulli$(q)$ random variables, $0 \leq q \leq 1$; and $\oplus$ represents modulo-2 addition.
By symmetry, we assume that $p_1 \leq 0.5$, $p_2 \leq 0.5$, and $q \leq 0.5$.

\subsubsection{Inner and Outer Bounds}
The following corollary gives an inner bound for the capacity
region of the binary noiseless MAC by applying a 
slightly generalized binary DPC at the informed encoder
in which the channel input $\rvx_1$ and the channel state $\rvs$ are correlated.

\begin{definition} \label{def:binary_innerbound}
Let $\calR^i(a_{10},a_{01})$ be the set of all rate pairs
 $(R_1,R_2)$ satisfying
\begin{subequations} \label{eqn:binary_innerbound}
\begin{align}
R_1 < & (1-q)\mbH_b(a_{10})+q\mbH_b(a_{01})  \\
R_2 < & \mbH_b(p_2)  \\
R_1+R_2 < & (1-q)\mbH_b(a_{10}) + q\mbH_b(a_{01}) \nonumber \\
              & +\mbH_b(p_2*[qa_{01}+(1-q)a_{10}]) \nonumber \\
             & -\mbH_b(qa_{01}+(1-q)a_{10}),
\end{align}
\end{subequations}
for $(a_{10},a_{01}) \in \calA$, where
\begin{equation*}
\calA := \{(x,y): 0 \leq x,~y \leq 1,~ \textrm{and}~ (1-q)x+q(1-y) \leq p_1\},
\end{equation*}
and $\mbH_b(\gamma):= -\gamma\log_2(\gamma)-(1-\gamma)\log_2(1-\gamma)$, and
$x*y := x(1-y)+y(1-x).$
Let
\begin{equation*}
\calR_{\mathrm{BIN}}^i := 
\mathrm{cl}\{\mathrm{co}\{\cup_{(a_{10},a_{01}) \in \calA}
\calR^i(a_{10},a_{01})\}\}.
\end{equation*}
\end{definition}

\begin{corollary}\label{cor:binary_innerbound}
The capacity region $\calC_{\mathrm{BIN}}$ for the binary noiseless state-dependent MAC with one informed encoder satisfies $\calR_{\mathrm{BIN}}^i \subseteq \calC_{\mathrm{BIN}}$.
\end{corollary}

\noindent \textbf{Proof:}  Encoding and decoding are similar to encoding and
decoding explained for the general DM case above.
The informed encoder uses  generalized
binary DPC  in which the random coding distribution allows arbitrary 
correlation between the channel input from the informed encoder and the known state.
We consider $\rvu_1=\rvx_1 \oplus \rvs$  and $\rvx_2 \sim \mathrm{Bernoulli}(p_2)$,
where: $\rvx_1$ is related to $\rvs$ by $a_{01}=P(\rvx_1=0|\rvs=1)$ and
 $a_{10}=P(\rvx_1=1|\rvs=0)$ with $a_{01}$ and $a_{10}$ chosen such
that $P(\rvx_1=1)\leq p_1$. We compute the region $\calR^{i}(a_{10},a_{01})$ defined in
 (\ref{thm:dm_innerbound}) using the
probability mass function of $\rvx_2$ and the auxiliary random variable $\rvu_1$ for
 all $(a_{10},a_{01}) \in \calA$, and deterministic $\rvq$
to obtain the region $\calR_{\mathrm{BIN}}^i$ in
(\ref{def:binary_innerbound}). We use deterministic $\rvq$ to compute the region in the binary case because we explicitly take the convex hull of unions of the regions computed with distributions. This completes the proof.

The following proposition provides a trivial outer bound for 
the capacity region of the binary noiseless MAC with one informed encoder.
We do not provide a proof because this outer bound can be easily obtained if we provide 
the channel state to the decoder.

\begin{proposition}\label{prop:binary_outerbound}
Let $\calR_{\mathrm{BIN}}^o$ be the set of all rate pairs $(R_1,R_2)$ satisfying
\begin{subequations} \label{eqn:binary_outerbound}
\begin{align}
R_1\leq & \mbH_b(p_1)  \\
R_2 \leq & \mbH_b(p_2)  \\
R_1+R_2 \leq &\left \{ \begin{array}{l l} 
\mbH_b(p_1+p_2) & \text{if $0 \leq p_1+p_2 < 0.5$} \\
1            & \text{if $0.5 \leq p_1+p_2 \leq 1$}
\end{array} \right .
\end{align}
\end{subequations}
 Then the capacity region $\calC_{\mathrm{BIN}}$ for the binary noiseless MAC
 with one informed encoder satisfies 
 $\calC_{\mathrm{BIN}} \subseteq \calR_{\mathrm{BIN}}^o$.
\end{proposition}

\subsubsection{Numerical Example}
\begin{figure}[ht]
\resizebox{1.0\columnwidth}{!}{
\includegraphics{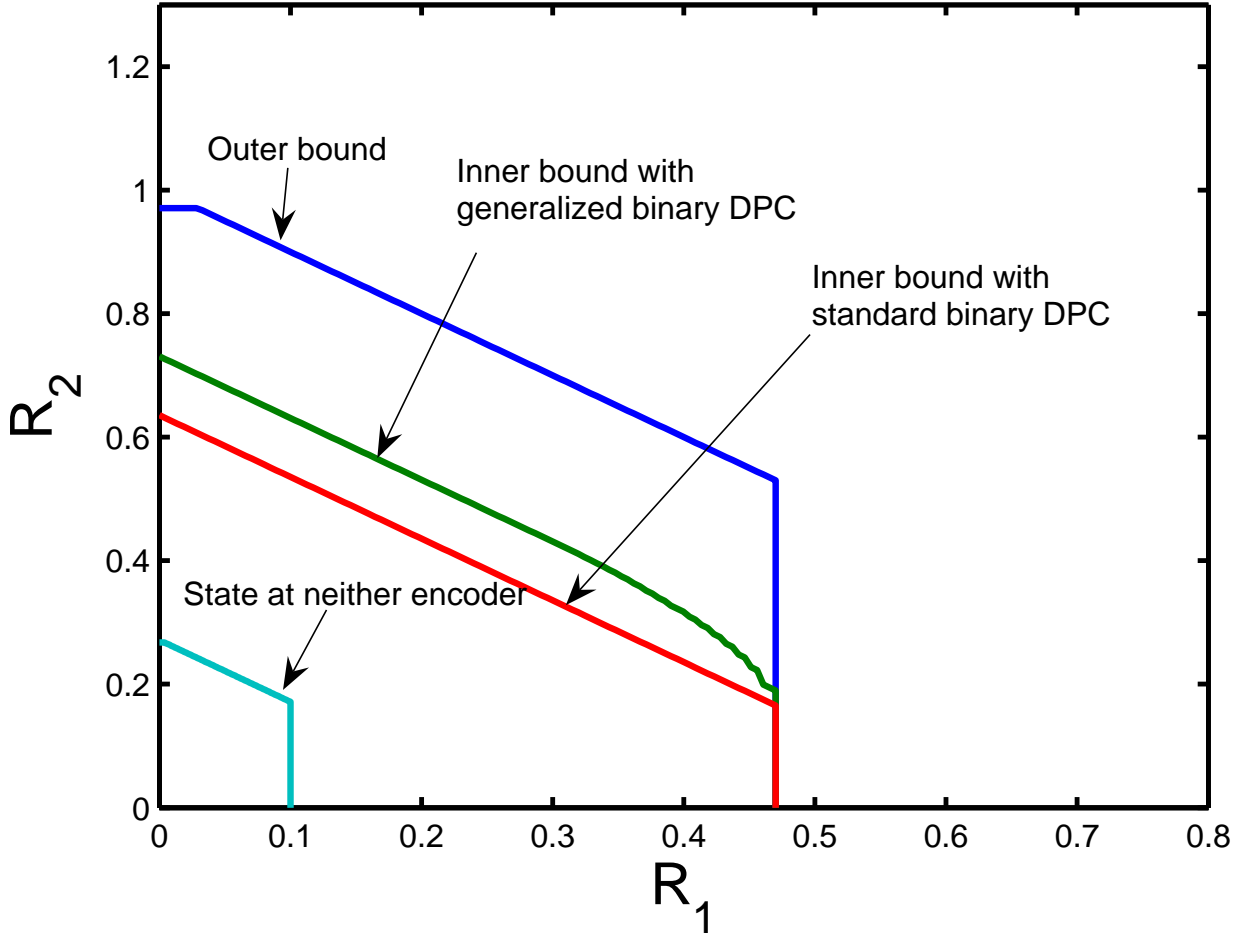}}
\caption{A numerical example of the binary noiseless multiple
access channel with $p_1=0.1$, $p_2=0.4$, $q=0.2$.} 
\label{fig:binary1}
\end{figure}

Figure \ref{fig:binary1} depicts the inner bound using 
generalized binary DPC specified in
Corollary~\ref{cor:binary_innerbound} and the outer bound specified in
Proposition~\ref{prop:binary_outerbound} for the case in which
$p_1=0.1$, $p_2=0.4$, and $q=0.2$.  Also shown for comparison are the
following: an inner bound using binary DPC alone, or the generalized
DPC with $a_{10}=p_1$ and $a_{01}=1-p_1$; and
the inner bound for the capacity region of the case in which the state is known to neither
the encoders nor the decoder.

These results show that the inner bound obtained by generalized
binary DPC is larger than that obtained using binary DPC \cite{zamir02:it}, and
suggest that the informed encoder can help the uninformed encoder 
using binary DPC \cite{zamir02:it} as well as
generalized binary DPC. Even though state is known to only one encoder,
both the encoders can benefit in terms of achievable rates compared to the case
in which state is unavailable at the encoder and the decoder. 

\subsubsection{Maximum Entropy State}
In this section, we discuss how the uninformed encoder benefits from the actions of the informed encoder 
even if $q=0.5$ so that $\mbH_b(\rvs)=1$.  The following corollary provides the capacity region of the
noiseless binary MAC with one informed encoder in this case.

\begin{corollary} \label{cor:bin_asym_capacity}
For the given input constraints $(p_1,p_2)$ and $q=0.5$, the capacity region of the binary noiseless MAC with one informed encoder is the set of rate pairs $(R_1,R_2)$ satisfying
\begin{subequations}\label{eqn:bin_asym_capacity}
\begin{align}
R_2  \leq &  \mbH_b(p_2)   \\
R_1+R_2  \leq & \mbH_b(p_1).
\end{align}
\end{subequations}
\end{corollary}

\noindent \textbf{Proof:} The region (\ref{cor:bin_asym_capacity}) is achieved if the informed encoder employs the generalized binary DPC with $a_{10}=p_1$ and $a_{01}=1-p_1$ or the standard binary DPC.  We obtain (\ref{eqn:bin_asym_capacity}) from (\ref{eqn:binary_innerbound}) by substituting $a_{10}=p_1$ and $a_{01}=1-p_1$ in (\ref{eqn:binary_innerbound}).
A converse proof for the above capacity region is given in Appendix~\ref{sec:conv_bin_asym_capacity}.

\noindent \textit{Remarks} 

\begin{itemize}
\item From (\ref{eqn:bin_asym_capacity}), we see that the uninformed encoder can achieve rates below $\min \{\mbH_b(p_1),\mbH_b(p_2) \}$ though the channel has maximum entropy state. Let us investigate how the uninformed encoder can benefit from the informed encoder's actions even in this case using successive decoding in which  $\rvu_1^n$ is decoded first using $\rvy^n$ and then $\rvx_2^n$ is decoded using $\rvy^n$ and $\rvu_1^n$. 
The informed encoder applies
the standard binary DPC, i.e., $a_{10}=p_1$ and $a_{01}=(1-p_1)$ in generalized binary
DPC, to generate its codewords, and the uninformed encoder uses a Bernoulli$(\tilde{p}_2)$ random
variable to generate its codewords, where $\tilde{p}_2 \leq p_2$. 
In the case of maximum entropy state, $\rvu_1^n$ can  be decoded first 
with arbitrary low probability of error if $R_1$ satisfies
\begin{equation}
R_1 < \mbH_b(p_1)- \mbH_b(\tilde{p}_2). \label{eqn:bin_R1_suc_RHS}
\end{equation}
for $\tilde{p}_2 \leq p_2$ and  $\tilde{p}_2 \leq p_1$.
The channel output can be written as
$\rvy_i = \rvu_{1,i} \oplus \rvx_{2,i}$ because 
$\rvu_{1,i} = \rvx_{1,i} \oplus \rvs_i$ for $i \in \{1,2,\ldots,n\}$.
Using $\rvu_1^n$,  we can generate a new channel
output for decoding $\rvx_2^n$ as
\[\tilde{\rvy}_i = \rvy_i \oplus \rvu_{1,i}  = \rvx_{2,i}\]
for $i \in \{1,2,\ldots,n\}$. Since there is no binary noise present in $\tilde{\rvy}^n$
for decoding $\rvx_2^n$,  the message of the
uninformed encoder can be decoded with
arbitrarily low probability of error if
$R_2 < \mbH_b(\tilde{p}_2)$ for $\tilde{p}_2$ 
satisfying  both $\tilde{p}_2 \leq p_1$ and $\tilde{p}_2 \leq p_2$.

Then the bound on $R_2$ can be written as
\[R_2 < \min \{\mbH_b(p_1),\mbH_b(p_2)\}.\]
If $p_1 > p_2$, we can achieve $R_2 < \mbH_b(p_2)$ as 
if there were no state in the channel, though the maximum entropy  
binary state is present in the channel and the state is not known to the uninformed encoder.
If $p_1 \leq p_2$, the uninformed encoder can still achieve positive rates, i.e., $R_2 <  \mbH_b(p_1)$.

\item Let us now discuss how the informed encoder achieves rate below $\mbH_b(p_1)$ using successive decoding in the reverse order, i.e., $\rvx_2^n$ is decoded first using
$\rvy^n$ and then $\rvu_1^n$ is decoded using $\rvy^n$ and $\rvx_2^n$.
If $q=0.5$, $\rvx_2^n$ can be decoded with arbitrary low
probability of error if $R_2 < [\mbH_b(p_2*(p_1*0.5))-\mbH_b(p_1*0.5)]=0$. This means that only
$R_2=0$ is achievable. Then $R_1 < \mbH_b(p_1)$ is achievable with
$a_{10}=p_1$ and $a_{01}=1-p_1$. 
\end{itemize} 

\begin{figure} [h]
\resizebox{1.0\columnwidth}{!}{
\includegraphics{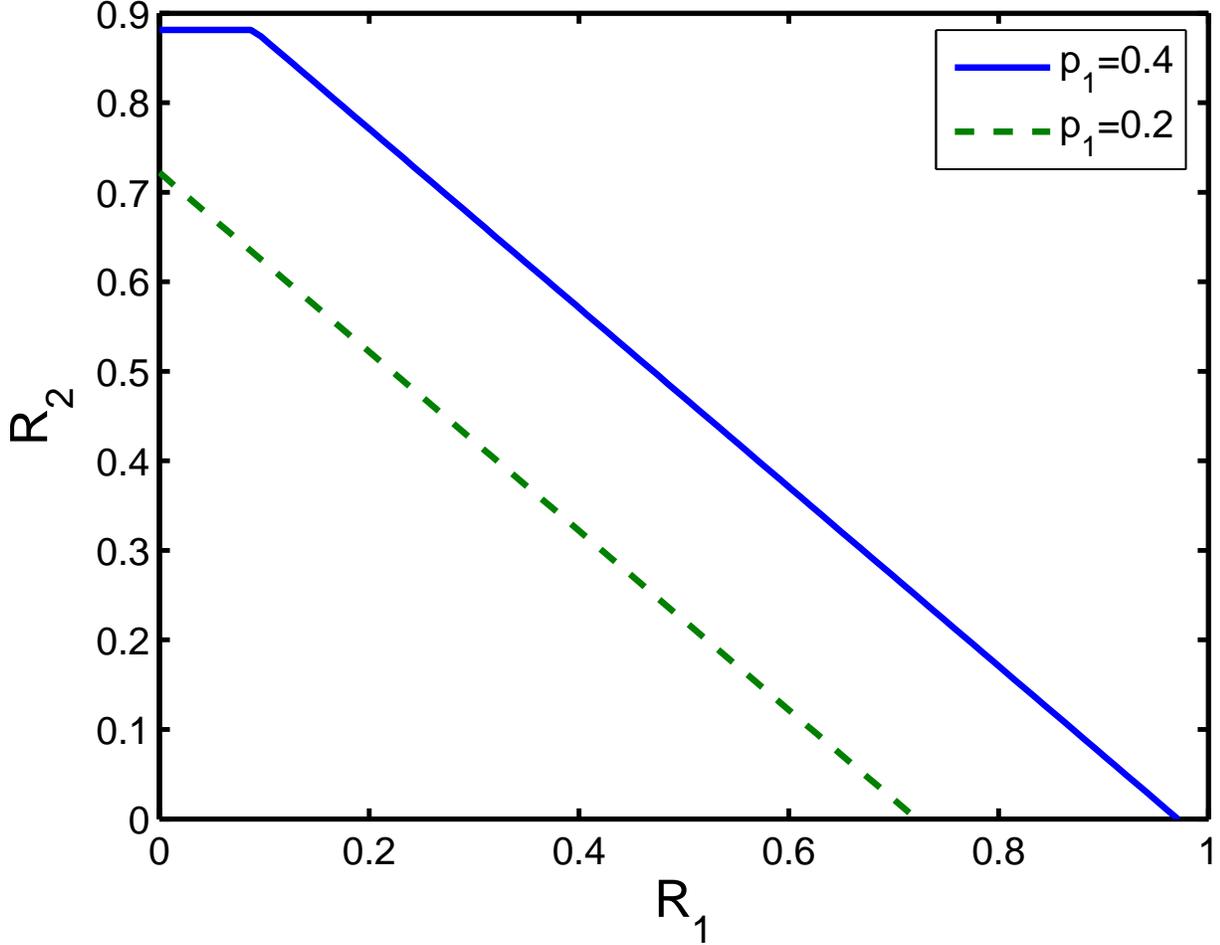}}
\caption{The capacity region of the binary noiseless MAC with maximum entropy binary state, i.e., $q=0.5$, and $p_2=0.3$.} \label{fig:binary_asymptotic}
\end{figure}

Let us illustrate the case of maximum entropy binary state 
with numerical examples.
Figure~\ref{fig:binary_asymptotic} illustrates the inner bound given in Corollary~\ref{cor:bin_asym_capacity} for $q=0.5$ and
$p_2=0.3$ in two cases $p_1=0.2$ ($p_1 \leq p_2 $) and $p_1=0.4$ ($p_1 > p_2$). In both cases, these numerical examples suggest that the uniformed encoder achieves positive rates from the actions of the informed encoder as discussed above.
In the case of $p_1=0.4$ ($p_1 > p_2$), the informed encoder can still achieve $\mbH_b(p_2)$ though the channel state has high entropy and is not known to the uninformed encoder, and the informed encoder has input constraint $p_1=0.4$. 

\section{Gaussian Memoryless Case}\label{sec:gaussiancase}
\begin{figure}[ht]
\psfrag{Informed Encoder}{Informed Encoder}
\psfrag{Uninformed Encoder}{Uninformed Encoder}
\psfrag{Decoder}{Decoder}
\psfrag{W1}{$\rvw_1$}
\psfrag{W2}{$\rvw_2$}
\psfrag{X1}{$\rvx_1^n$}
\psfrag{X2}{$\rvx_2^n$}
\psfrag{S}{$\rvs^n$}
\psfrag{Y}{$\rvy^n$}
\psfrag{Z}{$\rvz^n$}
\psfrag{D}{$(\hat{\rvw}_1,\hat{\rvw}_2)$}
\resizebox{.9\columnwidth}{!}{
\includegraphics{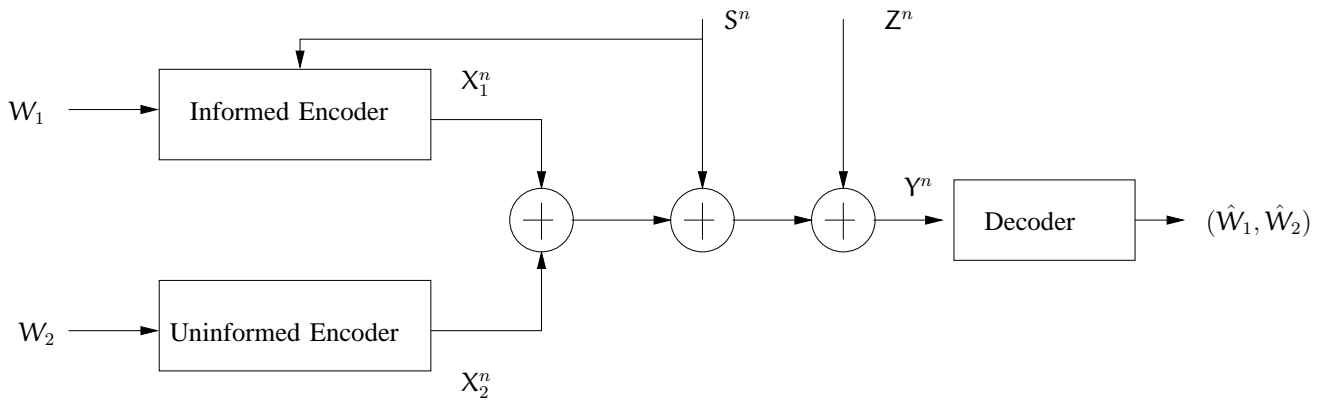}}
\caption{Gaussian state-dependent multiaccess channel with channel state
non-causally known to one encoder.} \label{fig:gmac}
\end{figure}

In this section, we develop inner and outer bounds for the memoryless Gaussian case.
The additive Gaussian MAC with one informed encoder is shown in Figure \ref{fig:gmac}.
The output of the channel is $\rvy^n=\rvx_{1}^{n}+\rvx_{2}^{n}+\rvs^{n}+\rvz^{n}$,
where: $\rvx_{1}^{n}$ and $\rvx_{2}^{n}$ are the channel inputs with
average power constraints $\sum_{i=1}^{n} \rvx_{1,i}^{2} \leq nP_1 $ and
$\sum_{i=1}^{n} \rvx_{2,i}^{2}
\leq nP_2$  with probability one, respectively;
$\rvs^n$ is the memoryless state vector whose elements are zero-mean 
Gaussian random variables with variance $Q$; and $\rvz^n$
is the memoryless additive noise vector whose elements are zero-mean
Gaussian random variables with variance $N$ and  independent of the
channel inputs and the state.

\subsection{Inner and Outer Bounds on the Capacity Region}
The following definition and theorem give an inner bound for the Gaussian MAC with
one informed encoder. To obtain the inner bound for this case, we apply  
generalized dirty paper coding (GDPC) at the informed encoder.

\begin{definition}
Let 
{\allowdisplaybreaks
\begin{subequations} \label{eqn:gaussian_RHSinnerbound}
\begin{align}
r_1(\rho,\alpha) &:= \frac{1}{2} \log
\left(\frac{P_1(1-\rho^2)(P_1+Q+2\rho \sqrt{P_1Q}+N)}
{P_1Q(1-\rho^2)(1-\alpha)^2 + N(P_1+\alpha^2Q +2\alpha\rho\sqrt{P_1Q})} \right )\\
r_2(\rho,\alpha) &:= \frac{1}{2} \log
\left(1+ \frac{P_2}{N+\frac{P_1Q(1-\rho^2)(1-\alpha)^2}
{(P_1+\alpha^2Q +2\alpha\rho\sqrt{P_1Q})}}
\right )\\
r_3(\rho,\alpha) &:=\frac{1}{2} \log
 \left(\frac{P_1(1-\rho^2)(P_1+P_2+Q+2\rho\sqrt{P_1Q}+N)}
{P_1Q(1-\rho^2)(1-\alpha)^2 + N(P_1+\alpha^2Q +2\alpha\rho\sqrt{P_1Q})} \right )
\end{align}
\end{subequations}}
for a given $-1\leq \rho \leq 0 $, and a given $\alpha \in \mathcal{A}(\rho)$, where
\[\mathcal{A}(\rho) = \{x \in \mathbb{R}: r_1(\rho,x)\geq 0, r_2(\rho,x)
\geq 0, r_3(\rho,x) \geq 0 \}.\]
\end{definition}

\begin{theorem} \label{thm:gaussian_innerbound}
Let $\calR^i(\rho,\alpha)$ be the set of all rate pairs $(R_1,R_2)$ satisfying
$R_1 < r_1(\rho,\alpha)$, $R_2 <  r_2(\rho,\alpha)$,
and $R_1+R_2 <  r_3(\rho,\alpha)$
for given $-1\leq \rho \leq 0 $ and $\alpha \in \mathcal{A}(\rho)$.
Let
\begin{equation}
\calR_{\mathrm{G}}^i = \mathrm{cl}\{\mathrm{co}\{\cup_{-1\leq \rho \leq 0,~ \alpha \in
 \mathcal{A}(\rho)} \calR^i(\rho,\alpha)\}\}.
\label{eqn:gaussianregion}
\end{equation}
Then the capacity region
$\calC_{\mathrm{G}}$ of the Gaussian MAC with one informed encoder
satisfies $\calR_{\mathrm{G}}^i \subseteq \calC_{\mathrm{G}}.$
\end{theorem}

\noindent \textbf{Proof:} Our results for the DM MAC can readily be extended
to memoryless channels with discrete time and continuous alphabets using
standard techniques \cite{gallager68:book}.
The informed encoder uses GDPC in which the random coding distribution allows 
arbitrary correlation between the channel input from the informed encoder and the known channel state.
Fix a correlation parameter $-1 \leq \rho \leq 0$. We then
consider the auxiliary random variable $\rvu_1= \rvx_{1}+\alpha \rvs$,
where $\alpha $ is a real number whose range will be discussed later,
$\rvx_{1}$ and $\rvs$ are correlated with correlation coefficient
$\rho$, $\rvx_1 \sim\mathcal{N}(0,P_1)$, and $\rvs \sim
\mathcal{N}(0,Q)$.  We consider $\rvx_2 \sim
\mathcal{N}(0,P_2)$. Encoding and decoding are performed similar to the
proof of Theorem~\ref{thm:dm_innerbound} in Section~\ref{proof_innerbound}. 
In this case, we assume that $q$ is deterministic because time sharing of 
regions with different distributions
is accomplished by explicitly taking convex hull of union of regions 
with different distributions. We
evaluate (\ref{eqn:dm_innerbound}) using the jointly Gaussian
distribution of random variables $\rvs$, $\rvu_1$, $\rvx_1$,
$\rvx_2$, $\rvz$, and $\rvy$ for a given $(\rho,\alpha)$ and obtain
$\calR^{i}(\rho,\alpha)$.  Also note that we restrict $\alpha$ to
$\mathcal{A}(\rho)= \{\alpha:\alpha\in \mathbb{R},r_1(\rho,\alpha)\geq
0,r_2(\rho,\alpha) \geq 0,r_3(\rho,\alpha) \geq 0\}$ for a given
$\rho$. By varying $\rho$ and $\alpha$, we obtain different achievable
rate regions $\mathcal{R}^{i}(\rho,\alpha)$. Taking the union of regions
$\mathcal{R}^i(\rho,\alpha)$ obtained by varying $\rho$ and $\alpha$
followed by taking the closure
and the convex hull operations completes the proof.

\noindent \textit{Remarks} 
\begin{itemize}
\item In both standard DPC \cite{mcosta83:it} and GDPC, the auxiliary random variable is given by $\rvu_1=\rvx_1+\alpha \rvs$. In GDPC, $\rvx_1 \sim \calN(0,P_1)$ and $\rvs \sim \calN(0,Q)$ are jointly correlated with correlation coefficient $\rho$, whereas in the standard DPC, they are uncorrelated. If the channel input $\rvx_1$ is negatively correlated with the channel state $\rvs$, then GDPC can be viewed as partial state cancellation followed by standard DPC. To see this, let us assume that $\rho$ is negative and denote $\hat{\rvx}_1$ as a linear estimate of $\rvx_1$ from $\alpha \rvs$ under the minimum mean square error (MMSE) criterion. Accordingly, $\hat{\rvx}_1=\alpha \rho \sqrt{\frac{P_1}{Q}}\rvs$. We can rewrite the auxiliary random variable $\rvu_1$ as follows
\begin{align}
\rvu_1 & = (\rvx_1- \hat{\rvx}_1)+ \hat{\rvx}_1+\alpha \rvs \nonumber \\
       & =  \rvx_{1,w} + \alpha \left(1+\rho \sqrt{\frac{P_1}{Q}}\right) \rvs \nonumber \\
       & =  \rvx_{1,w} + \alpha \left(1-\sqrt{\frac{\gamma P_1}{Q}}\right ) \rvs  \nonumber \\
       &=   \rvx_{1,w} + \alpha \hat{\rvs}  
\end{align} 
where $\gamma = \rho^2 \in (0,1]$, $\hat{\rvs}$ can be viewed as the remaining state after state cancellation using power $\gamma P_1$, and $\rvx_{1,w}$ is error with variance $(1-\gamma) P_1$ and is uncorrelated with $\hat{\rvs}$.
GDPC with negative correlation coefficient $\rho$ can be interpreted as  standard DPC with power $(1-\gamma)P_1$ applied on the remaining state $\hat{\rvs}$ after state cancellation using power $\gamma P_1$.

\item In this paper, we focus on the two-encoder model in which one is informed and the other is uninformed, but the concepts can be extended to the model with any number of uninformed and informed encoders. The informed encoders apply  GDPC to help the uninformed encoders. Following \cite{gelfand83:pisit, kim04:pisit}, the informed encoders cannot be affected from the actions of the other informed encoders because the informed encoders can eliminate the effect of the remaining state on their transmission after the state cancellation by them.
\end{itemize}

The following proposition gives a trivial outer bound for the capacity
region of the Gaussian MAC with  one informed encoder.
We do not provide a proof because this bound is 
the capacity region of the additive white Gaussian MAC with \textit{all}
informed encoders \cite{gelfand83:pisit, kim04:pisit}, the  capacity region
of the additive white Gaussian MAC with state
known to only the decoder, and the capacity region of the additive white Gaussian
MAC without state. 

\begin{proposition}\label{prop:gaussian_outerbound}
Let $\calR_{\mathrm{G}}^o$ be the set of all rate pairs $(R_1,R_2)$ satisfying
{\allowdisplaybreaks
\begin{subequations} \label{eqn:gaussian_outerbound}
\begin{align}
R_1\leq & \frac{1}{2}\log \left (1+ \frac{P_1}{N}\right )   \\
R_2 \leq & \frac{1}{2}\log\left (1+ \frac{P_2}{N}\right )  \\
R_1+R_2 \leq & \frac{1}{2}\log\left (1+ \frac{P_1+P_2}{N}\right ).
\end{align}
\end{subequations}}
Then the capacity region $\calC_{\mathrm{G}}$ for the
Gaussian MAC with one informed encoder satisfies
$\calC_{\mathrm{G}} \subseteq \calR_{\mathrm{G}}^o$.
\end{proposition}

\subsection{Numerical Example}
\begin{figure} [h]
\resizebox{1.0\columnwidth}{!}{
\includegraphics{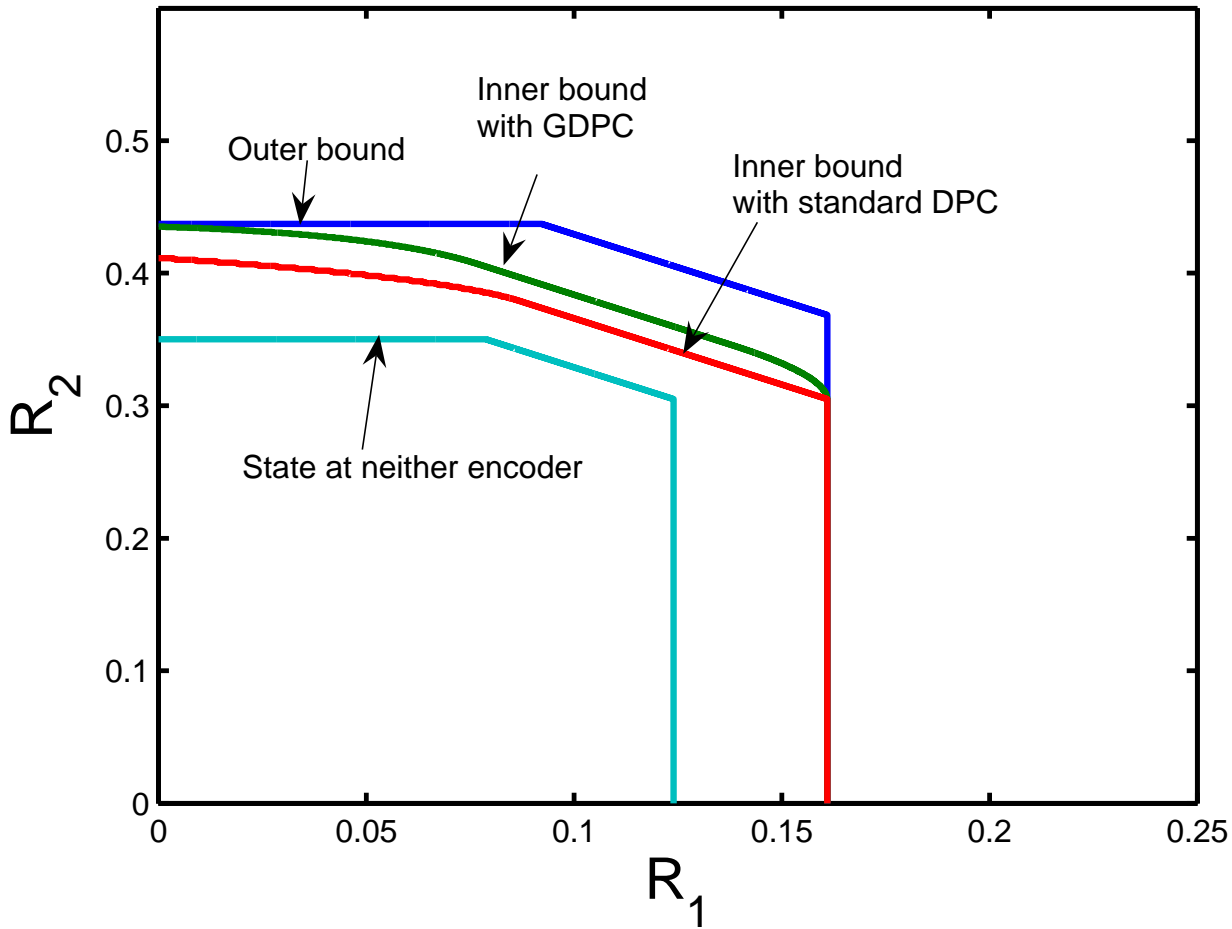}
}
\caption{An achievable region for Gaussian MAC with
$P_1=15$, $P_2=50$, $Q=20$, and $N=60$.} \label{fig:achievable}
\end{figure}

Figure \ref{fig:achievable} depicts the inner bound using  GDPC given
in Theorem~\ref{thm:gaussian_innerbound} and the outer bound specified
in Proposition~\ref{prop:gaussian_outerbound} for the case in which
$P_1=15$, $P_2=50$, $Q=20$, and $N=60$. Also shown for comparison are the
following: an inner bound using DPC alone, or GDPC with $\rho=0$ and $\alpha$ as 
parameter; and the capacity region for the case in which the
 the state is unavailable at the encoders and the decoder.

These results suggest that the informed encoder
can help the uninformed encoder using DPC as well as GDPC.
Even though the state is known only at one encoder,
both the encoders benefit from this situation by allowing negative
correlation between the channel input $\rvx_1$ and the state $\rvs$
at the informed encoder, since the negative correlation allows the informed
encoder to partially cancel the state. The achievable rate region
$\mathcal{R}^{i}(0,\alpha)$ obtained by
applying DPC \cite{mcosta83:it}  with $\alpha$ as a parameter
is always contained in
$\mathcal{R}_{\mathrm{G}}^i$ in (\ref{eqn:gaussianregion}).
In contrast  to the case of  state available to both the encoders
\cite{gelfand83:pisit, kim04:pisit}, GDPC is not sufficient to completely mitigate
the effect of state on the capacity region.

\begin{figure} [h]
\psfrag{R2}{$R_{2,\mathrm{max}}$}
\resizebox{1.0\columnwidth}{!}{
\includegraphics{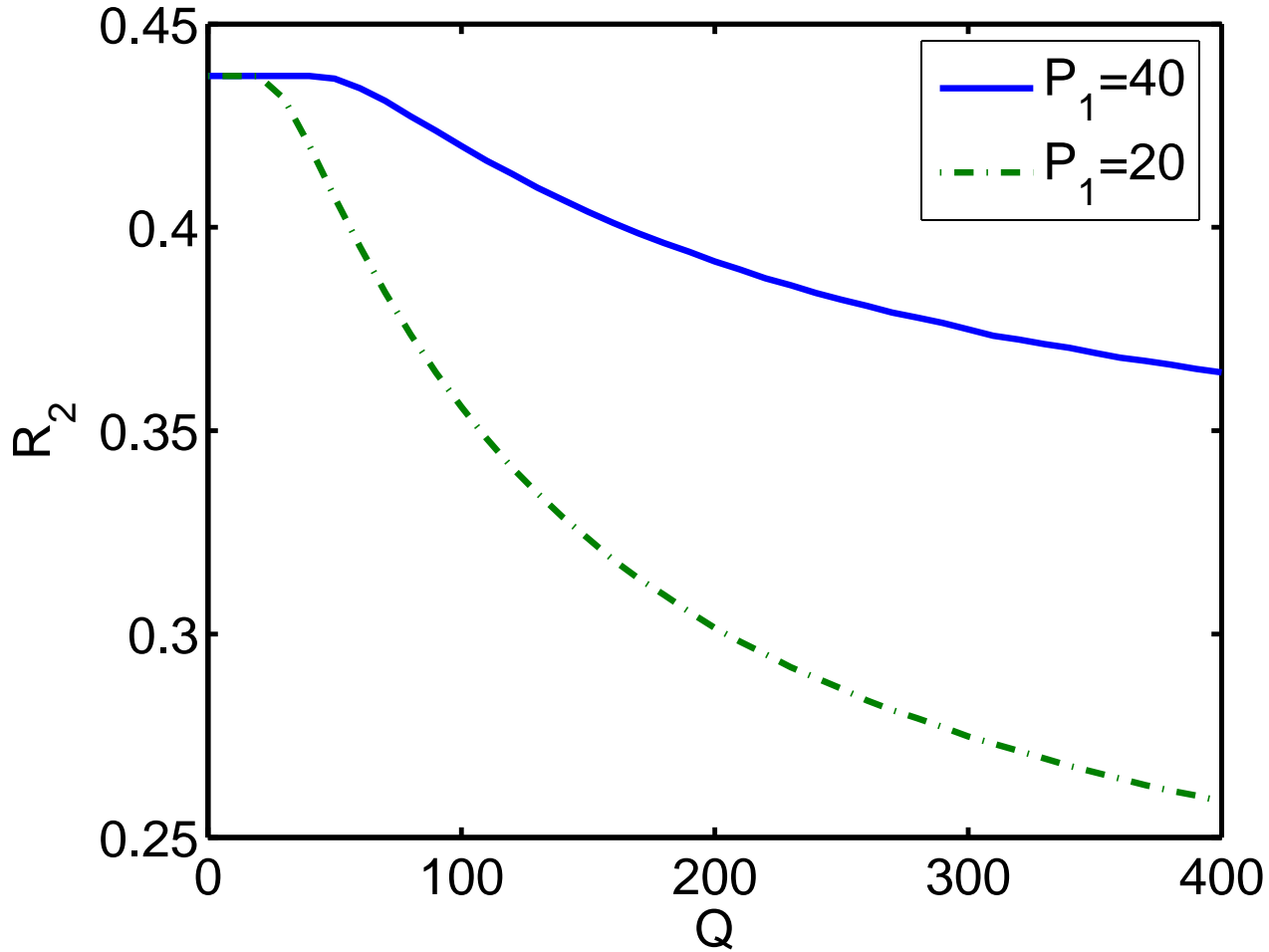}}
\caption{Variation of the maximum rate of uninformed encoder $R_{2,\max}$ 
with the channel state variance $Q$ when $R_1=0$, $P_2=50$,  and $N=60$.} \label{fig:R2VSQ}
\end{figure}

Figure~\ref{fig:R2VSQ} illustrates how the maximum rate of the uninformed encoder $R_{2,\mathrm{max}}$ 
varies with the channel state variance $Q$ if $R_1=0$, for $P_2=50$, and $N=60$. As shown in Figure~\ref{fig:R2VSQ}, $R_{2,\mathrm{max}}$ decreases as $Q$ increases because the variance of remaining state also increases following state cancellation by the informed encoder. The decrease in $R_{2,\mathrm{max}}$ is slower as $P_1$ increases because the informed encoder can help the uninformed encoder more in terms of achievable rates as its power increases.

\subsection{Asymptotic Analysis}
In this section, we discuss the inner bound in  Theorem~\ref{thm:gaussian_innerbound}
 as $Q \rightarrow \infty$.
\begin{definition}
Let $\tilde{\calR}^i(\rho,\alpha)$ be the set of all rate pairs $(R_1,R_2)$ satisfying
$R_1 <  \tilde{r}_1(\rho,\alpha)$,
 $R_2 <  \tilde{r}_2(\rho,\alpha)$,
 and $R_1+R_2 <  \tilde{r}_3(\rho,\alpha)$
for a given $-1\leq \rho \leq 0 $ and
$\alpha \in \tilde{\calA}(\rho)=
\{x \in \mathbb{R}: 0 \leq x \leq \frac{2P_1(1-\rho^2)}{P_1(1-\rho^2)+N} \}$,
where \footnote{$r_i(\rho,\alpha)$ for $i=1,2,3$ is defined in (\ref{eqn:gaussian_RHSinnerbound}) and is function of $Q$, though variable $Q$ is not mentioned in the notation $r_i(\rho,\alpha)$.}
{\allowdisplaybreaks
\begin{subequations} \label{eqn:gaussian_RHSinnerbound1}
\begin{align}
\tilde{r}_1(\rho,\alpha)=\lim_{Q \rightarrow \infty} r_1(\rho,\alpha)  &= \frac{1}{2}
\log \left(\frac{P_1(1-\rho^2)}{P_1(1-\rho^2)(1-\alpha)^2 + \alpha^2 N} \right )
 \label{eqn:R1_gaussian_RHSinnerbound1}\\
 \tilde{r}_2(\rho,\alpha)=
\lim_{Q \rightarrow \infty} r_2(\rho,\alpha) &=
\frac{1}{2}
 \log \left(1+ \frac{P_2}{N+ \frac{P_1(1-\rho^2)(1-\alpha)^2}{\alpha^2}} \right )
\label{eqn:R2_gaussian_RHSinnerbound1}\\
\tilde{r}_3(\rho,\alpha) =
\lim_{Q \rightarrow \infty} r_3(\rho,\alpha) &=\frac{1}{2}
\log \left(\frac{P_1(1-\rho^2)}
{P_1(1-\rho^2)(1-\alpha)^2 + \alpha^2 N} \right )
 \label{eqn:R3_gaussian_RHSinnerbound1}.
\end{align}
\end{subequations}}
\end{definition}

\begin{corollary}\label{cor:gaussian_asym_innerbound}
As the variance of the state becomes very large, i.e., $Q \rightarrow \infty$, an inner bound for the capacity region
of the Gaussian MAC with one informed encoder is given by
\[\tilde{\calR}_{\mathrm{G}}^i = \mathrm{cl}\{\mathrm{co}\{\cup_{-1\leq \rho \leq 0,~ \alpha \in
 \mathcal{A}(\rho)} \tilde{\calR}^i(\rho,\alpha)\}\}.\]
\end{corollary}

\noindent \textit{Remarks:}
\begin{itemize}
\item Let us investigate how the uninformed encoder can benefit from the
informed encoder's  actions even as $Q \rightarrow \infty$. For this discussion,
consider successive decoding  in which the auxiliary codeword $\rvu_1^n$ of the informed encoder is
decoded first using  the channel output $\rvy^n$ and then the codeword $\rvx_2^n$ of
 the uninformed encoder is
 decoded using $\rvy^n$ and $\rvu_1^n$.
 In the limit as $Q\rightarrow \infty$, $\rvu_1^n$ can
 be decoded first with arbitrary low probability of error if
 $R_1$ satisfies
\begin{equation}
R_1 < \frac{1}{2} \log \left (\frac{P_1(1-\rho^2)}
{P_1(1-\rho^2)(1-\alpha)^2 + \alpha^2 (P_2+N)} \right ), \label{eqn:R1_suc_RHS}
\end{equation}
where $\rho \in [-1,0]$ and
$0 \leq \alpha \leq \frac{2P_1(1-\rho^2)}{P_1(1-\rho^2)+P_2+N}$.
The right hand side of (\ref{eqn:R1_suc_RHS}) is obtained by calculating the
expression $\mbI(\rvu_1;\rvy)-\mbI(\rvu_1,\rvs)$ for the assumed jointly
Gaussian distribution and letting
$Q \rightarrow \infty$.
The channel output can be written as
$\rvy_i = \rvu_{1,i} + \rvx_{2,i} + (1-\alpha)\rvs_{i}+\rvz_i $ because 
$\rvu_{1,i} = \rvx_{1,i}+ \alpha \rvs_i$
for $i \in \{1,2,\ldots,n\}$.
The estimate of  $(1-\alpha)\rvs_i$ using $\rvu_{1,i}$ is denoted as $\hat{\rvs}_i$ for
$i \in \{1,2,\ldots,n\}$.

Using $\hat{\rvs}^n$ and $\rvu_1^n$,  we can generate a new channel
output for decoding $\rvx_2^n$ as
\[\tilde{\rvy}_i = \rvy_i -\rvu_{1,i} - \hat{\rvs}_{i} = \rvx_{2,i}+ \rvz_i + 
((1-\alpha)\rvs_i -\hat{\rvs}_{i} )\]
for $i \in \{1,2,\ldots,n\}$. Since all random variables are identical, we omit the
subscript $i$ for further discussion.
The variance of  total noise present in elements of $\tilde{\rvy}^n$ for decoding
$\rvx_2^n$ is $N+\frac{P_1(1-\rho^2)(1-\alpha)^2}{\alpha^2}$, where $N$ is
the variance of $\rvz$, and $\frac{P_1(1-\rho^2)(1-\alpha)^2}{\alpha^2}$ is the error of
estimating $(1-\alpha)\rvs$ from $\rvu_1$. Then the message of the
uninformed encoder can be decoded with
arbitrarily low probability of error if
$R_2 < \lim_{Q \rightarrow \infty} r_2(\rho,\alpha)$ for given $\rho \in [-1,0]$ and
$0 \leq \alpha \leq \frac{2P_1(1-\rho^2)}{P_1(1-\rho^2)+P_2+N}$. Even if the variance of
the state becomes infinite, nonzero rate for
 the uninformed encoder can be achieved  because the estimation error is finite
  for $\rho \in [-1,0]$ due
to the increase of the variance of $\rvu_1$ with the increase of the state variance.

Our aim is to minimize the variance of the estimation error
$((1-\alpha)\rvs -\hat{\rvs})$ to
maximize $r_2(\rho, \alpha)$ over $\rho$ and $\alpha$.
Since the right hand side of (\ref{eqn:R1_suc_RHS}) becomes non-negative for
$0 \leq \alpha \leq \frac{2P_1(1-\rho^2)}{P_1(1-\rho^2)+P_2+N}$ and
$\rho \in [-1,0]$, we consider only these values.
The variance of the estimation error is decreasing in both
 $\rho \in [-1,0]$ and $\alpha \in [0,1]$ and
 is increasing in the remaining range of $\alpha$.
Then  $r_2(\rho, \alpha)$ achieves its maximum at
$\rho =0$ and $\alpha =\min\{1,\frac{2P_1}{P_1+P_2+N}\}$.
If $P_1 \geq P_2+N$, so that $R_1$ is nonnegative, then
\[R_2 < \frac{1}{2}\log\left(1+ \frac{P_2}{N} \right)\]
is achievable.
In this case, the uninformed encoder fully benefits from actions of the informed encoder, specifically from its auxiliary codewords, even though the variance of interfering state is very large.
If $P_1 < P_2+N$, then $R_2 < \lim_{Q \rightarrow \infty}r_2(0,\alpha^*)$
is achievable
where $\alpha^* = \frac{2P_1}{P_1+P_2+N}$. In either case,
 GDPC with $\rho=0$  is optimal in terms of
assisting the uninformed encoder,
contrary to the case of finite state variance.
This makes sense because, if the state has infinite variance, then it is impossible
for the informed
encoder to explicitly cancel it with finite power.
 
\item To investigate how the informed encoder achieves its maximum rate, let us
consider successive decoding in the reverse order in which
 $\rvx_2^n$ is decoded first using
$\rvy^n$ and then $\rvu_1^n$ is decoded using $\rvy^n$ and $\rvx_2^n$.
As $Q \rightarrow \infty$, $\rvx_2^n$ can be decoded with arbitrary low
probability of error if
$R_2 < \lim_{Q \rightarrow \infty} \mbI(\rvx_2,\rvy)=0$. This means that only
 $R_2=0$ is achievable.
Then $R_1 < \frac{1}{2}\log\left(1+\frac{P_1}{N} \right)$ is achievable with
$\rho =0$ and $\alpha = \frac{P_1}{P_1+N}$.
\end{itemize}

The following proposition gives an outer bound for the Gaussian MAC with one informed encoder as 
$Q \rightarrow \infty.$ 

\begin{proposition}\label{prop:gaussian_asym_outerbound}
As $Q \rightarrow \infty$, an outer bound for the capacity region of the 
Gaussian MAC with one informed encoder is the
set of rate pairs $(R_1,R_2)$ satisfying
\begin{subequations} \label{eqn:gaussian_asym_outerbound}
\begin{align}
R_2 \leq & \frac{1}{2}\log\left (1+ \frac{P_2}{N}\right )  \\
R_1+R_2 \leq & \frac{1}{2}\log\left (1+ \frac{P_1}{N}\right ).
\end{align}
\end{subequations}
\end{proposition}

We do not provide a proof of the above proposition because the proof is similar to the converse proof given in Appendix~\ref{sec:conv_bin_asym_capacity}. The outer bound in Proposition~\ref{prop:gaussian_asym_outerbound} is better than the trivial outer bound in Proposition~\ref{prop:gaussian_outerbound} obtained by giving the channel state to the decoder. 

\begin{figure} 
\resizebox{1.0\columnwidth}{!}{
\includegraphics{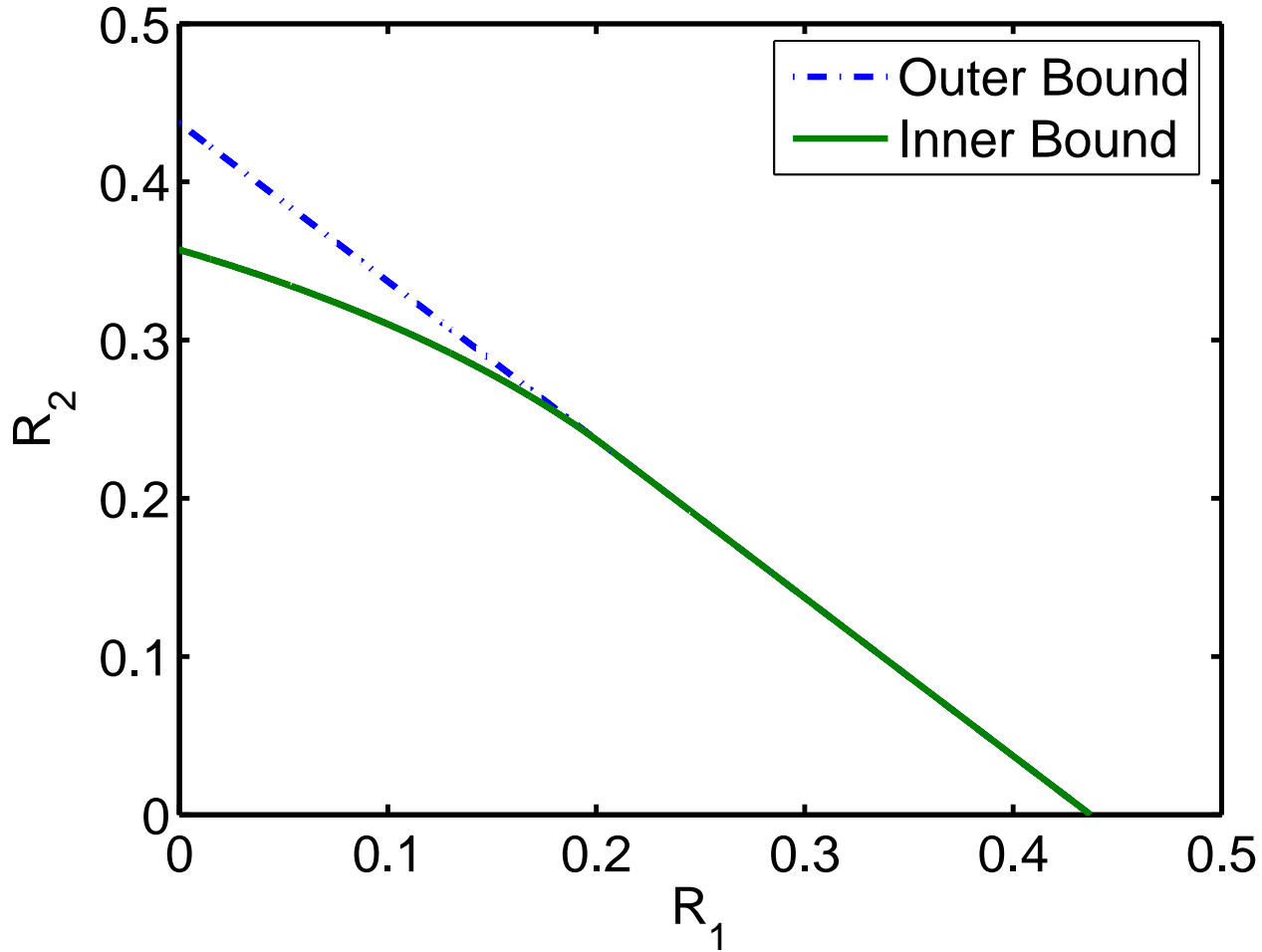}}
\caption{The inner and outer bounds for the capacity region of the Gaussian MAC with one informed encoder in the strong additive Gaussian state case, i.e., $Q \rightarrow \infty$, for $P_1=50$ ($P_1 \leq P_2+N$), $P_2=50$ and $N=60$.} \label{fig:Gauss_asymptotic_P150}
\end{figure}

\begin{figure} 
\resizebox{1.0\columnwidth}{!}{
\includegraphics{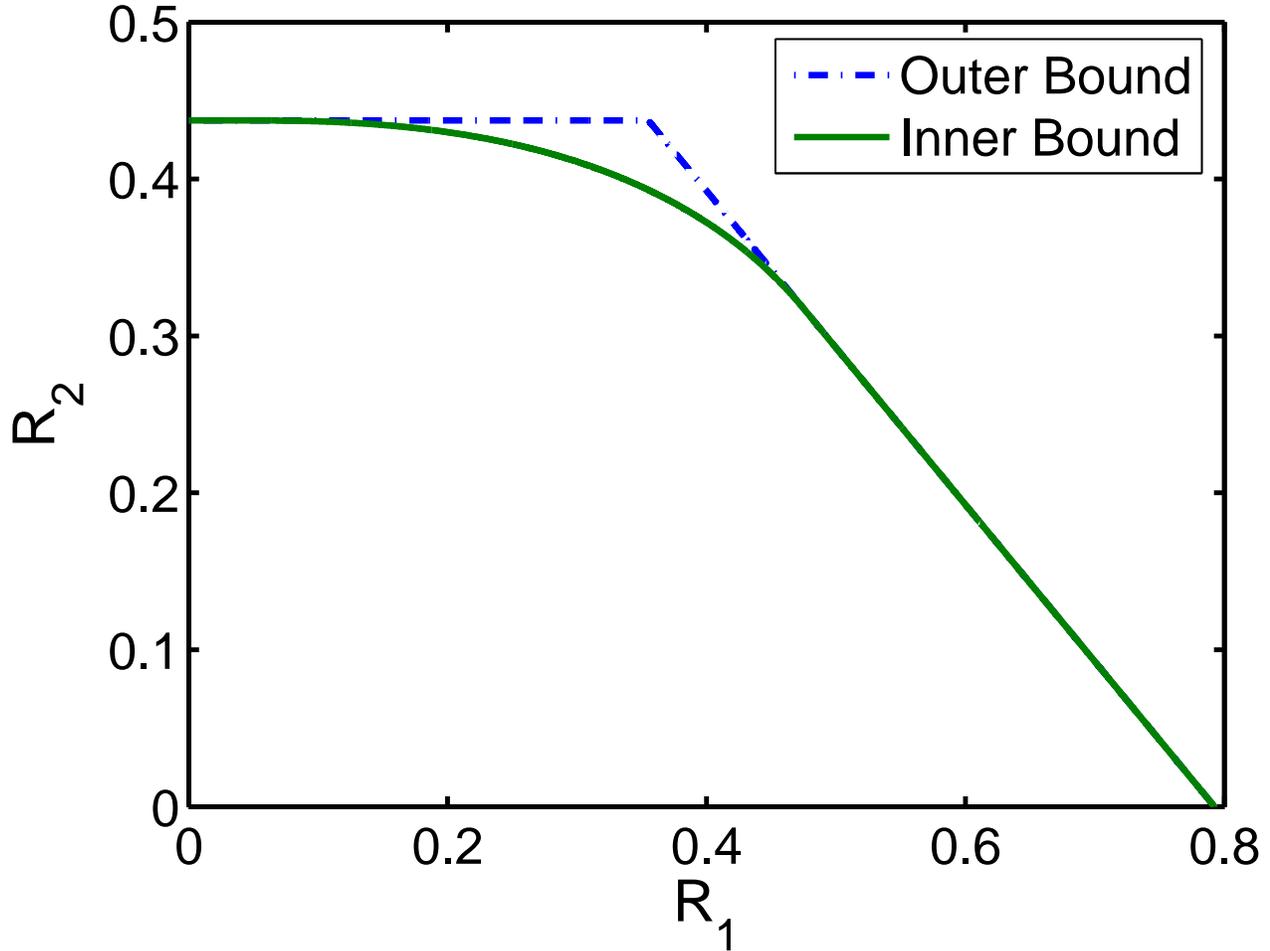}}
\caption{The inner and outer bounds for the capacity region of the Gaussian MAC with one informed encoder in the strong additive Gaussian state case, i.e., $Q \rightarrow \infty$, for $P_1=120$ ($P_1 >P_2+N$), $P_2=50$ and $N=60$.} \label{fig:Gauss_asymptotic_P1120}
\end{figure}

Finally, let us discuss the case of strong additive Gaussian channel state, 
i.e., $Q \rightarrow \infty$, with  numerical examples.
Figure~\ref{fig:Gauss_asymptotic_P150} and Figure~\ref{fig:Gauss_asymptotic_P1120} illustrate the inner bound in Corollary~\ref{cor:gaussian_asym_innerbound} and the outer bound in Proposition~\ref{prop:gaussian_asym_outerbound}  in two cases, $P_1=50$ ($P_1 \leq P_2 +N$) and $P_1=120$ ($P_1 > P_2 +N$), respectively, for $P_2=50$ and $N=60$.  In both cases, the uniformed encoder achieves positive rates from the actions of the informed encoder as discussed above. In the case of $P_1=120$ ($P_1 > P_2 +N$), the informed encoder  
can still achieve $\frac{1}{2} \log(1+\frac{P_2}{N})$, though the additive channel state is 
very strong and is not known to the uninformed encoder, and the informed encoder has finite  power.

\begin{figure}
\resizebox{1.0\columnwidth}{!}{
\includegraphics{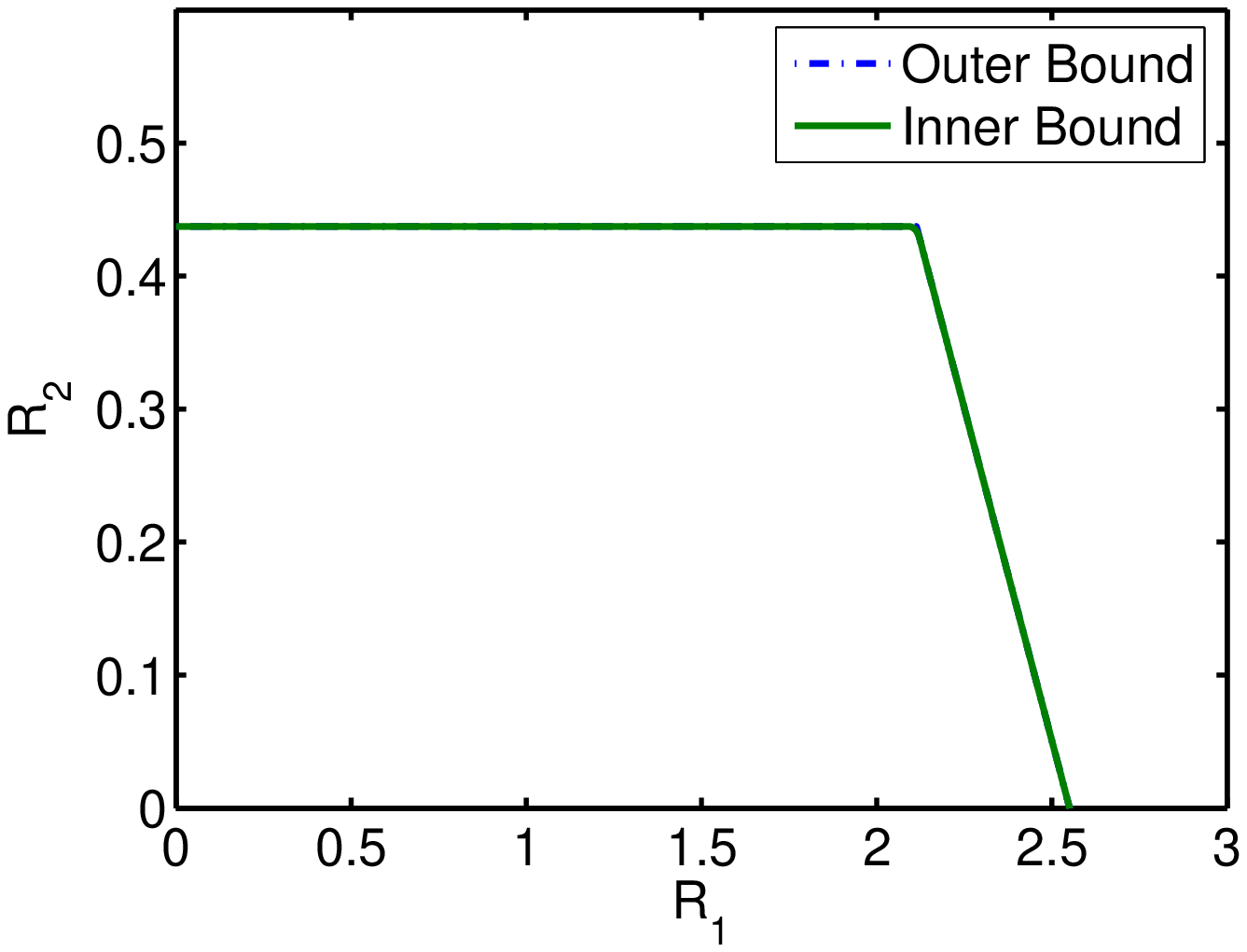}}
\caption{The inner and outer bounds for the capacity region of the Gaussian MAC with one informed encoder in the strong additive Gaussian state case, i.e., $Q \rightarrow \infty$, for $P_1=2000$,$P_2=50$ and $N=60$.} \label{fig:Gauss_asymptotic_P12000}
\end{figure}

As $P_1$ increases and $P_1 \geq P_2+N$, in the strong additive Gaussian state case, the inner bound in Corollary~\ref{cor:gaussian_asym_innerbound} and the outer bound in Proposition~\ref{prop:gaussian_asym_outerbound}
meet asymptotically. Thus, we obtain the capacity region for $P_1 \rightarrow \infty$ and $P_1 \geq P_2+N$. 
For the very large values of $P_1$, the outer bound given in Proposition~\ref{prop:gaussian_asym_outerbound} is achieved asymptotically with $\alpha=1.$ Figure~\ref{fig:Gauss_asymptotic_P12000} shows the inner bound in the strong additive state case which is also compared with the outer bound in Proposition~\ref{prop:gaussian_asym_outerbound} for the very large values of $P_1$, i.e., $P_1=2000$.

\section{Conclusions}\label{sec:conclusions}
In this paper, we considered a state-dependent MAC with state known to some, but not all, encoders.
We derived an inner bound for the DM case and specialized to a noiseless binary case using generalized binary DPC.
If the channel state is a Bernoulli$(q)$ random variable with $q<0.5$, we compared the inner bound in the binary case with a trivial outer bound obtained by providing the channel state to only the decoder. The inner bound obtained by generalized binary DPC does not meet the trivial outer bound for $q<0.5$. For $q=0.5$, we obtain the capacity region for binary noiseless case by deriving a non-trivial outer bound.  

For the Gaussian case, we also derived an inner bound using GDPC and an outer bound by providing the channel state to the decoder also. It appears that the uninformed encoder benefits from GDPC because explicit state cancellation is present in GDPC. In the case of strong Gaussian state, i.e., the variance of state going to infinity, we also specialized the inner bound and analyzed how the uninformed encoder benefits from the auxiliary codewords of the informed encoder even in this  case because explicit state cancellation is not helpful for this case. In the case of strong channel state, we also derived a non-trivial outer bound which is tighter than the trivial outer bound. These bounds asymptotically meet if $P_1 \rightarrow \infty$ and $P_1 \geq P_2+N$. From results in the special cases of both the binary case and the Gaussian case, we note that the inner bounds meet the non-trivial outer bounds. From the results and observations in this paper, we would like to conclude that we are not able to show that random coding techniques and inner bounds in this paper achieve the capacity region due to lack of non-trivial outer bounds in all cases for this problem.

\appendix
We denote the set of jointly strongly typical
sequences~\cite{cover91:book, csiszar81:book} with distribution
$p(\svx,\svy)$ as $T_{\epsilon}^n[\rvx,\rvy]$.  We define
$T_{\epsilon}^n[\rvx,\rvy|\svx^n]$ as the following
\[T_{\epsilon}^n[\rvx,\rvy|\svx^n]=\{\svy^n:(\svx^n,\svy^n)\in T_{\epsilon}^n[\rvx,\rvy]\}.\]

\subsection{Proof of Theorem~\ref{thm:dm_innerbound}} \label{proof_innerbound}

In this section, we construct a sequence of codes
$(\lceil 2^{nR_1}\rceil,\lceil2^{nR_2}\rceil,n)$
with $P_e^n \rightarrow 0$ as
$n \rightarrow \infty$ if $(R_1,R_2)$ satisfies Equation~(\ref{eqn:dm_innerbound}).
The random coding used in this section is a combination of 
Gel'fand-Pinsker coding \cite{gelfand80:pcit} and coding for MAC \cite{cover91:book}.
This random coding is not a new technique but it is included for completeness.
Fix $\epsilon > 0$ and take 
$(\rvq,\rvs,\rvu_1,\rvx_1,\rvx_2,\rvy)\in \calP^i$.
\subsubsection{Encoding and Decoding}
\begin{itemize}
\item \textbf{Encoding:}
The encoding strategy at the two encoders is as follows.
Let $M_{1}=2^{n(R_1-4\epsilon)}$, $M_2=2^{n(R_2-2\epsilon)}$,
 and $J=2^{n(\mathbb{I}(\rvu_{1};\rvs|\rvq)+2\epsilon)}$.
At the informed encoder, where the state is available,
generate $JM_1$ sequences $\rvu_1^{n}(\svq^n,m_1,j)$,
 whose elements are drawn i.i.d. with $ p(\svu_1|\svq)$, for each time sharing random
  sequence $\rvq^n$,  where $1\leq m_1 \leq M_1$, and $1\leq j \leq J$.
Here, $m_1$ indexes bins and $j$ indexes sequences within
a particular bin $m_1$. For encoding, given state $\rvs^n=\svs^n$, time sharing sequence
$\rvq^n=\svq^n$  and message $\rvw_{1}\in \{1,2,\dots,M_1\}$, look in
bin $\rvw_{1}$ for a sequence $\rvu^{n}_{1}(\svq^n,\rvw_{1},j), 1\leq j \leq J$,
such that  $\rvu^{n}_{1}(\svq^n,\rvw_{1},j) \in T_{\epsilon}^n[\rvq,\rvu,\rvs|\svq^n,\svs^n]$.
Then the informed encoder generates $\rvx_{1}^{n}$ from $(\rvu_1^n,\rvs^n,\rvq^n)$ according to
probability law $\prod_{j=1}^n p(\svx_{1,j}|\svu_{1,j},\svs_j,\svq_j)$.

At the uninformed encoder, sequences $\rvx_{2}^{n}(\svq^n,m_2)$, whose elements are drawn
i.i.d. with $p(\svx_2|\svq) $, are generated for each time sharing sequence
$\rvq^n=\svq^n$, where $1 \leq m_{2} \leq M_2$. The uninformed encoder chooses
 $\rvx_{2}^{n}(\svq^n,\rvw_2)$ to send the
message $\rvw_{2}\in \{1,2,\dots,M_{2}\}$ for a given time-sharing sequence
$\rvq^n =\svq^n$ and sends the codeword $\rvx_{2}^n$.

Given the inputs and the state, the decoder receives $\rvy^n$
according to conditional probability distribution $\prod_{i} p(\svy_i|\svs_i,\svx_{1,i},\svx_{2,i})$.
It is assumed that the time-sharing sequence $\rvq^n =\svq^n$ is non-causally
known to both the encoders and the decoder.

\item \textbf{Decoding:}
The decoder, upon receiving the sequence $\rvy^n$, chooses a pair
$(\rvu^{n}_{1}(\svq^n,m_1,j),\rvx_{2}^{n}(m_2))$,
$1\leq m_1 \leq M_1$, $1\leq j \leq J$, and $ 1 \leq m_2 \leq M_2$ such that
$(\rvu_1^n(\svq^n,m_1,j),\rvx_2^n(\svq^n,m_2)) \in
 T_{\epsilon}^n[\rvq,\rvu_1,\rvx_2,\rvy|\svq^n,\rvy^n]$.
If such a pair exists and is unique, the decoder declares that
 $(\hat{\rvw}_{1},\hat{\rvw}_{2})=(m_1, m_2)$. Otherwise, the decoder declares an error.
 \end{itemize}

\subsubsection{Analysis of Probability of Error}
The average probability of error is given by {\allowdisplaybreaks
\begin{align}
P_e^n =& \sum_{\svs^n\in \calS^n,\svq^n \in \calQ^n}
p(\svs^n)p(\svq^n)\mathrm{Pr}[\mathrm{error}|\svs^n,\svq^n] \nonumber \\
 \leq & \sum_{\svs^n\not\in T_{\epsilon}^n[\rvs]}p(\svs^n)+
 \sum_{\svq^n\not\in T_{\epsilon}^n[\rvq]}p(\svq^n) \nonumber \\
 &+\sum_{\svs^n\in T_{\epsilon}^n[\rvs],
 \svq^n\in T_{\epsilon}^n[\rvq]}p(\svq^n)\mathrm{Pr}[\mathrm{error}|\svs^n,\svq^n].
  \label{eqn:asym_prob_error}
\end{align}}
The first term, $\mathrm{Pr}[\svs^n\not\in T_{\epsilon}^n[\rvs]]$, and the
second term, $\mathrm{Pr}[\svq^n\not\in T_{\epsilon}^n[\rvq]]$,
in the right hand side expression of (\ref{eqn:asym_prob_error})
go to zero as $n \rightarrow \infty$ by
the strong asymptotic equipartition property (AEP) \cite{cover91:book}.

Without loss of generality, we can assume that $(\rvw_1,\rvw_2)=(1,1)$ is sent,
time sharing sequence is $\rvq^n=\svq^n$, and state realization is $\rvs^n =\svs^n$.
The probability of error is given by the conditional probability of
error given $(\rvw_1,\rvw_2)=(1,1)$,
$\rvq^n=\svq^n \in T_{\epsilon}^n [\rvq]$, and
$\rvs^n=\svs^n \in T_{\epsilon}^n [\rvs] $.

\begin{itemize} 
\item Let $E_1$ be the event that there is no sequence $\rvu_1^n(\svq^n,\rvw_1,j)$
such that  $\rvu_1^n(\svq^n,1,j) \in T_{\epsilon}^n[\rvq,\rvu_1,\rvs|\svq^n,\svs^n]$.
For any $\rvu_{1}^n(\svq^n,1,j)$ and $\rvs^{n}=\svs^n$  generated independently according
to $\prod p(\svu_{1i}|\svq_i)$ and $\prod p(\svs_{i})$, respectively, the probability that
there exists at least one $j$ such that $\rvu_{1}^{n}(\svq^n,1,j)
\in  T_{\epsilon}^n[\rvq,\rvu,\rvs|\svq^n,\svs^n]$
 is greater than $(1-\epsilon)2^{-n(\mathbb{I}(\rvu_{1};\rvs|\rvq)+\epsilon)}$ for
$n$ sufficiently large.
There are  $J$ number of such $\rvu_{1}^{n}$'s in each bin.
The probability of event $E_1$, the probability that there is no
$\rvu_{1}^{n}$ for a given $\svs^n$ in a particular bin, is  therefore bounded by
\begin{align}
\mathrm{Pr} [E_1] &\leq
[1-(1-\epsilon)2^{-n(\mathbb{I}(\rvu_{1};\rvs|\rvq)+\epsilon)}]^
{2^{n(\mathbb{I}(\rvu_{1};\rvs|\rvq)+2\epsilon)}}.   \label{above1}
\end{align}
Taking the natural logarithm on both sides of (\ref{above1}), we obtain
\begin{align}
\ln(\mathrm{Pr}[E_1]) &\leq
2^{n(\mathbb{I}(\rvu_{1};\rvs|\rvq)+2\epsilon)}\ln[1-(1-\epsilon)
2^{-n(\mathbb{I}(\rvu_{1};\rvs|\rvq)+\epsilon)}]  \nonumber \\
&\stackrel{(a)}{\leq}  -2^{n(\mathbb{I}(\rvu_{1};\rvs|\rvq)+2\epsilon)}
(1-\epsilon)2^{-n(\mathbb{I}(\rvu_{1};\rvs|\rvq)+\epsilon)}  \nonumber \\
&=                    -(1-\epsilon)2^{n\epsilon}, \label{eqn:prob_E1}
\end{align}
where $(a)$ follows from the inequality $\ln(q)\leq (q-1)$.
From (\ref{eqn:prob_E1}), $\mathrm{Pr}[E_1] \rightarrow 0$ as
$n \rightarrow \infty$.

Under the event $E_1^c$, we can also assume that a particular
sequence $\rvu_{1}^{n}(\svq^n,1,1)$ in bin 1 is jointly strongly
typical with $\rvs^n=\svs^n$.  Thus, codewords $\rvx_{1}^{n}$
corresponding to the pair $(\rvu_1^{n}(\svq^n,1,1),\svs^n)$ and $\rvx_{2}^{n}$
corresponding to $\rvx_{2}^n(\svq^n,1)$ are
sent from the informed and the uninformed encoders,
respectively.

\item Let $E_2$ be the event that
$$(\rvu^{n}_{1}(\svq^n,1,1),\rvx_{2}^{n}(\svq^n,1),\rvy^n) \not \in
T_{\epsilon}^n[\rvq,\rvu_1,\rvx_2,\rvy|\svq^n].$$
The Markov lemma \cite{cover91:book}  ensures jointly strong typicality of 
$(\svq^n,\svs^n,\rvu_{1}^{n}(\svq^n,1,1),\rvx^{n}_{2}(\svq^n,1),,\rvy^n)$
 with high probability if $(\svq^n,\svs^n,\rvu_{1}^{n}(\svq^n,1,1),\rvx_{1}^n)$ is
jointly strongly typical and $(\svq^n,\rvx^{n}_{2}(1))$ is jointly
strongly typical. We can conclude that  $\mathrm{Pr}[E_2|E_{1}^{c}] \rightarrow 0$
as $n \rightarrow \infty$.

\item Let $E_3$ be the event that 
$$\rvu^n_1(\svq^n,m_1,j) \in
 T_{\epsilon}^n[\rvq,\rvu_1,\rvx_2,\rvy|\svq^n,\rvy^n,\rvx_2^n(\svq^n,1)].$$
The probability that $\rvu^n_1(\svq^n,m_1,j) \in
 T_{\epsilon}^n[\rvq,\rvu_1,\rvx_2,\rvy|\svq^n,\rvy^n,\rvx_2^n(\svq^n,1)]$
for ($m_1=1$ and $j \neq 1$), or ($m_1\neq 1$ and $1\leq j \leq J$),
 is less than $2^{-n(\mathbb{I}(\rvu_1;\rvy|\rvx_2,\rvq)-\epsilon)}$  for
sufficiently large $n$.
There are approximately $JM_1$ (exactly $JM_1-1$) such
$\rvu_1^n$ sequences in the codebook.
Thus, the  conditional probability of event
$E_3$ given $E_{1}^{c}$ and $E_{2}^{c}$ is upper bounded by
\begin{align}
\mathrm{Pr}[E_3|E_{1}^{c},E_{2}^{c}] &\leq 
 2^{-n((\mathbb{I}(\rvu_1;\rvy|\rvx_2,\rvq)-\mathbb{I}(\rvu_1;\rvs|\rvq))-R_1) +\epsilon)}.
\label{eqn:prob_E3}
\end{align}
From (\ref{eqn:prob_E3}),  $\mathrm{Pr}[E_3|E_{1}^{c},E_2^c] \rightarrow 0$
as $n \rightarrow \infty$
if $R_1 <\mathbb{I}(\rvu_1;\rvy|\rvx_2,\rvq)-\mathbb{I}(\rvu_1;\rvs|\rvq)$ and $\epsilon > 0$.

\item Let $E_4$ be the event that 
$$\rvx_{2}^{n}(\svq^n,m_2) \in
T_{\epsilon}^n[\rvq,\rvu_1,\rvx_2,\rvy|\svq^n,\rvy^n,\rvu_1^n(\svq^n,1,1)]$$
 for $m_2\neq 1$.
The probability that $\rvx_{2}^{n}(\svq^n,m_2) \in
T_{\epsilon}^n[\rvq,\rvu_1,\rvx_2,\rvy|\svq^n,\rvy^n,\rvu_1^n(\svq^n,1,1)]$
for $m_2\neq 1$ is less than
$2^{-n(\mathbb{I}(\rvx_2;\rvy|\rvu_1,\rvq)-\epsilon)}$ for sufficiently large $n$.
There are approximately $M_2=2^{n(R_2-2\epsilon)}$ such
$\rvx_{2}^{n}$ sequences in the codebook. Thus, the  conditional
probability of event
$E_4$ given $E_{1}^{c}$ and $E_{2}^{c}$ is upper bounded by
\begin{align}
\mathrm{Pr}[E_4|E_{1}^{c},E_{2}^{c}] &\leq
2^{-n(\mathbb{I}(\rvx_2;\rvy|\rvu_1,\rvq)-R_2 +\epsilon)}.
\label{eqn:prob_E4}
\end{align}
From (\ref{eqn:prob_E4}), $\mathrm{Pr}[E_4|E_{1}^{c},E_{2}^{c}]
\rightarrow 0$ as $n \rightarrow \infty$
if $R_2 < \mathbb{I}(\rvx_2;\rvy|\rvu_1,\rvq)$.

\item 
Finally, let $E_5$ be the event that 
$$(\rvu^{n}_{1}(\svq^n,m_1,j),\rvx_{2}^{n}(\svq^n,m_2))
\in T_{\epsilon}^n[\rvq,\rvu_1,\rvx_2,\rvy|\svq^n,\rvy^n]$$ for
(($m_1=1$ and $j \neq 1$), or ($m_1\neq 1$ and $1\leq j \leq J$)), and $m_2 \neq 1$.
The probability that $(\rvu^{n}_{1}(\svq^n,m_1,j),\rvx_{2}^{n}(\svq^n,m_2)) \in
T_{\epsilon}^n[\rvq,\rvu_1,\rvx_2,\rvy|\svq^n,\rvy^n]$ for $m_1\neq 1$, $1\leq j \leq J$,
and $m_2 \neq 1$ is less than
$2^{-n(\mathbb{I}(\rvu_1,\rvx_2;\rvy|\rvq)-\epsilon)}$, for sufficiently large $n$.
There are approximately $JM_1$ sequences $\rvu_1^n$ and $M_2$  sequences $\rvu_{2}^{n}$
in the codebook. Thus, the  conditional probability of event
$E_5$ given $E_{1}^{c}$ and $E_{2}^{c}$ is upper bounded by
\begin{align}
\mathrm{Pr}[E_5|E_{1}^{c},E_{2}^{c}] &\leq
 2^{-n((\mathbb{I}(\rvu_1,\rvx_2;\rvy|\rvq)-\mathbb{I}(\rvu_1;\rvs|\rvq))-(R_1+R_2) +3\epsilon)}.
 \label{eqn:prob_E5}
\end{align}
From (\ref{eqn:prob_E5}), the $\mathrm{Pr}[E_5|E_{1}^{c},E_{2}^{c}]
\rightarrow 0$ as $n \rightarrow \infty$
if $R_2+R_2 < \mathbb{I}(\rvu_1,\rvx_2;\rvy|\rvq)-\mathbb{I}(\rvu_1;\rvs|\rvq)$.
\end{itemize}

In terms of these events, $\mathrm{Pr}[\mathrm{error}|\svs^n,\svq^n]$ in
 (\ref{eqn:asym_prob_error}) can be
upper-bounded via the union bound, and
 the fact that probabilities are less than one, as
\begin{align}
\mathrm{Pr}[\mathrm{error}|\svs^n,\svq^n] \leq & \mathrm{Pr}[E_1]+ \mathrm{Pr}[E_2| E_{1}^{c}]
 + \mathrm{Pr} [E_3|E_{1}^{c},E_{2}^{c}]\nonumber \\
 &+ \mathrm{Pr} [E_4|E_{1}^{c},E_{2}^{c}]+ \mathrm{Pr} [E_5|E_{1}^{c},E_{2}^{c}].
\label{eqn:asym_unionbound}
\end{align}

From (\ref{eqn:asym_unionbound}), it can be easily seen that
 $\mathrm{Pr}[\mathrm{error}|\svs^n,\svq^n]
\rightarrow 0$ as $n \rightarrow \infty$.
Therefore, the probability of error $P_e^n$ goes to zero as $n \rightarrow \infty$ from (\ref{eqn:asym_prob_error})
and completes the proof.

\subsection{Converse for the Capacity Region in Corollary~\ref{cor:bin_asym_capacity}} \label{sec:conv_bin_asym_capacity}
In this section, we show that $(R_1,R_2)$ satisfies (\ref{eqn:bin_asym_capacity}) for any given sequence 
of binary codes $(\lceil 2^{nR_1} \rceil, \lceil 2^{nR_2} \rceil, n)$ for the noiseless binary state-dependent
MAC with $q=0.5$ and one informed encoder satisfying $\lim_{n \rightarrow \infty} P_e^n =0$. 

Let us first bound the rate of the uninformed encoder as follows.
{\allowdisplaybreaks
\begin{align}
nR_2 & \leq \mbH(\rvw_2) \nonumber \\
    & = \mbH(\rvw_2|\rvw_1,\rvs^n) \nonumber \\
    & \stackrel{(a)}{\leq} \mbI(\rvw_2;\rvy^n|\rvw_1,\rvs^n)+n\epsilon_n \nonumber \\
    & \stackrel{(b)}{\leq} \sum_{j=1}^n \mbI(\rvx_{2,j};\rvy_j|\rvx_{1,j},\rvs_j)+n\epsilon_n \nonumber \\
    & = \sum_{j=1}^n \mbH(\rvx_{2,j})+n\epsilon_n \nonumber \\ 
    &\stackrel{(c)}{=}  \sum_{j=1}^n \mbH_b(p_{2,j})+n\epsilon_n \nonumber \\ 
    &\stackrel{(d)}{\leq} n \mbH_b\left(\frac{1}{n}\sum_{j=1}^np_{2,j}\right)+n\epsilon_n \nonumber \\ 
    &\stackrel{(e)}{\leq}  n\mbH_b\left(p_2\right)+n\epsilon_n, \label{eqn:conv_temp1}  
\end{align}}
where: \\
$(a)$ follows from Fano's inequality and $\epsilon_n \rightarrow 0$ as $P_e^n \rightarrow 0$, \\
$(b)$ follows from the fact that $\rvx_1^n$ and $\rvx_2^n$ are deterministic functions of $(\rvw_1,\rvs^n)$ and $(\rvw_2)$, respectively, the memoryless property of the channel, and $\mbH(\rvy_j|\rvx_1^n,\rvs^n,\rvw_1) \leq \mbH(\rvy_j|\rvx_{1,j},\rvs_j),$  \\
$(c)$ follows from the fact that $\rvx_{2,j}$ is a Bernoulli$(p_{2,j})$ 
satisfying $\frac{1}{n}\sum_{j=1}^n p_{2,j} \leq p_2$,\\
$(d)$ follows from the fact that the binary entropy function is a concave function,\\
$(e)$ follows from the fact that the binary entropy function is a monotone increasing function 
in the interval between $0$ and $0.5$, and $\frac{1}{n}\sum_{j=1}^n p_{2,j} \leq p_2 \leq 0.5$.

Let us bound $R_1+R_2$ as follows.
{\allowdisplaybreaks
\begin{align}
n(R_1+R_2)  \leq & \mbH(\rvw_1,\rvw_2) \nonumber \\
            \stackrel{(a)}{=}&\mbI(\rvw_1,\rvw_2:\rvy^n)+n\epsilon_n \nonumber \\
            =& \mbH(\rvy^n)-\mbH(\rvy^n|\rvw_1,\rvw_2,\rvs^n) \nonumber \\
            &+\mbH(\rvy^n|\rvw_1,\rvw_2,\rvs^n)-\mbH(\rvy^n|\rvw_1,\rvw_2)+n\epsilon_n \nonumber \\
           =&\mbI(\rvw_1,\rvw_2,\rvs^n;\rvy^n)-\mbI(\rvs^n;\rvy^n|\rvw_1,\rvw_2)+n\epsilon_n \nonumber \\
           \stackrel{(b)}{=}& \mbI(\rvx_1^n,\rvx_2^n,\rvs^n;\rvy^n) 
          -\mbI(\rvs^n;\rvy^n|\rvw_1,\rvw_2,\rvx_2^n)+n\epsilon_n\nonumber \\
          = &\sum_{j=1}^n [\mbH(\rvy_j|\rvy^{j-1}) - \mbH(\rvy_j|\rvy^{j-1},\rvx_1^n,\rvx_2^n,\rvs^n) \nonumber \\
           & - \mbH(\rvs_j|\rvw_1,\rvw_2,\rvx_2^n,\rvs^{j-1}) \nonumber \\
           &+ \mbH(\rvs_j|\rvw_1,\rvw_2,\rvx_2^n,\rvs^{j-1},\rvy^n)]+n\epsilon_n\nonumber \\
           \stackrel{(c)}{\leq}& \sum_{j=1}^n [\mbH(\rvy_j) - \mbH(\rvy_j|\rvx_{1,j},\rvx_{2,j},\rvs_j) \nonumber \\
            &- \mbH(\rvs_j) + \mbH(\rvs_j|\rvx_{2,j},\rvy_j)]+n\epsilon_n \nonumber \\
           =& \sum_{j=1}^n [\mbI(\rvx_{1,j},\rvx_{2,j},\rvs_j;\rvy_j) 
             - \mbI(\rvx_{2,j},\rvy_j;\rvs_j)]+n\epsilon_n \nonumber \\
          \stackrel{(d)}{=} &   \sum_{j=1}^n [ \mbI(\rvx_{2,j};\rvy_j)+\mbI(\rvs_j;\rvy_j|\rvx_{2,j}) \nonumber \\
          &+\mbI(\rvx_{1,j};\rvy_j|\rvx_{2,j},\rvs_j)- \mbI(\rvs_j;\rvy_j|\rvx_{2,j})]+n \epsilon_n \nonumber \\
           =&\sum_{j=1}^n [\mbI(\rvx_{2,j};\rvy_j)+\mbI(\rvx_{1,j};\rvy_j|\rvx_{2,j},\rvs_j)]+n \epsilon_n \nonumber \\
           \stackrel{(e)}{=}&\sum_{j=1}^n [\mbI(\rvx_{2,j};\rvy_j)+\mbH(\rvx_{1,j}|\rvs_j)]+n \epsilon_n \nonumber \\
           \stackrel{(f)}{=}&\sum_{j=1}^n [\mbH_b(p_{2,j}*(0.5(a_{10,j}+a_{01,j}))) \nonumber \\
           &- \mbH_b(0.5(a_{10,j}+a_{01,j})) \nonumber \\
           &+0.5\mbH_b(a_{10,j})+0.5\mbH_b(a_{01,j})]+n \epsilon_n \nonumber \\
           \stackrel{(g)}{\leq} & \sum_{j=1}^n \mbH_b(p_{1,j}) + n \epsilon_n \nonumber \\
           \stackrel{(h)}{\leq} & n \mbH_b \left( \frac{1}{n }\sum_{j=1}^n p_{1,j}\right) + n \epsilon_n \nonumber \\
            \stackrel{(i)}{\leq} & n \mbH_b \left(p_1 \right) + n \epsilon_n, \label{eqn:conv_temp2}        
\end{align}}
where: \\
$(a)$ follows from Fano's inequality and $\epsilon_n \rightarrow 0$ as $P_e^n \rightarrow 0$, \\
$(b)$ follows from the fact that $\rvx_1^n$ and $\rvx_2^n$ are deterministic functions of $(\rvw_1,\rvs^n)$ and $\rvw_2$, respectively, and the memoryless property of the channel \\
$(c)$ follows from the fact that $\mbH(\rvy_j|\rvy^{j-1}) \leq \mbH(\rvy_j)$, 
$\mbH(\rvy_j|\rvy^{j-1},\rvx_1^n,\rvx_2^n,\rvs^n)=\mbH(\rvy_j|\rvx_{1,j},\rvx_{2,j},\rvs_j),$ \\
$\mbH(\rvs_j|\rvw_1,\rvw_2,\rvx_2^n)=\mbH(\rvs_j)$, and $\mbH(\rvs_j|\rvw_1,\rvw_2,\rvx_2^n,\rvy^n)\leq \mbH(\rvs_j|\rvx_{2,j},\rvy_j),$ \\
$(d)$ follows from the fact that $\mbI(\rvs_j;\rvx_{2,j})=0,$ \\
$(e)$ follows from the fact that $\mbH(\rvy_j|\rvx_{2,j},\rvs_j)=\mbH(\rvx_{1,j}|\rvs_j)$ and $\mbH(\rvy_j|\rvx_{1,j},\rvx_{2,j},\rvs_j)=0,$ \\
$(f)$ follows from the fact that the $\rvx_{2,j}$ is a Bernoulli$(p_{2,j})$ random variable with 
$\sum_{j=1}^n p_{2,j} \leq np_2$; and $\rvx_{1,j}$ is correlated to $\rvs_j$ with $a_{10,j}=\mathrm{Pr}(\rvx_{1,j}=1|\rvs_j=0)$ and $a_{01,j}=\mathrm{Pr}(\rvx_{1,j}=0|\rvs_j=1)$ satisfying
\begin{align}
\mathrm{Pr}(\rvx_{1,j}=1)=p_{1,j}=& \mathrm{Pr}(\rvs_j=1)\mathrm{Pr}(\rvx_{1,j}=1|\rvs_j=1) \nonumber \\
&+\mathrm{Pr}(\rvs_j=0)\mathrm{Pr}(\rvx_{1,j}=1|\rvs_j=0) \nonumber \\ 
&= 0.5(a_{10,j}+(1-a_{01,j})), \nonumber
\end{align}
$(g)$ follows from the fact that the term $[\mbH_b(p_{2,j}*(0.5(a_{10,j}+a_{01,j}))) - \mbH_b(0.5(a_{10,j}+a_{01,j})) +0.5\mbH_b(a_{10,j})+0.5\mbH_b(a_{01,j})]$ is maximized under the constraint $0.5(a_{10,j}+(1-a_{01,j}))=p_{1,j}$ for values $a_{10,j}=p_{1,j}$ and $a_{01,j}=1-p_{1,j},$ and the maximum value of the term $[\mbH_b(p_{2,j}*(0.5(a_{10,j}+a_{01,j}))) - \mbH_b(0.5(a_{10,j}+a_{01,j})) +0.5\mbH_b(a_{10,j})+0.5\mbH_b(a_{01,j})]$ is $\mbH(p_{1,j})$ \\
$(h)$ follows from the concavity property of the binary entropy function, \\
$(i)$ follows from the fact that the binary entropy function is a monotone increasing function 
in the interval between $0$ and $0.5$, and $\frac{1}{n}\sum_{j=1}^n p_{1,j} \leq p_1 \leq 0.5$.

From (\ref{eqn:conv_temp1}) and (\ref{eqn:conv_temp2}), we can conclude that the rate pair $(R_1,R_2)$ satisfies 
(\ref{eqn:bin_asym_capacity}) by letting $n$ go to $\infty.$

\bibliographystyle{IEEEtran}
\bibliography{shiva}

\begin{thebibliography}{10}
\providecommand{\url}[1]{#1}
\csname url@rmstyle\endcsname
\providecommand{\newblock}{\relax}
\providecommand{\bibinfo}[2]{#2}
\providecommand\BIBentrySTDinterwordspacing{\spaceskip=0pt\relax}
\providecommand\BIBentryALTinterwordstretchfactor{4}
\providecommand\BIBentryALTinterwordspacing{\spaceskip=\fontdimen2\font plus
\BIBentryALTinterwordstretchfactor\fontdimen3\font minus
  \fontdimen4\font\relax}
\providecommand\BIBforeignlanguage[2]{{%
\expandafter\ifx\csname l@#1\endcsname\relax
\typeout{** WARNING: IEEEtran.bst: No hyphenation pattern has been}%
\typeout{** loaded for the language `#1'. Using the pattern for}%
\typeout{** the default language instead.}%
\else
\language=\csname l@#1\endcsname
\fi
#2}}

\bibitem{swanson98:picc}
M.~D. Swanson, M.~Kobayashi, and A.~H. Tewfik, ``{Multimedia Data-Embedding and
  Watermarking Technologies},'' in \emph{\textit{Proc. {IEEE} Int. Conf.
  Communications (ICC)}}, vol.~2, 1998, pp. 823--827.

\bibitem{bchen00}
B.~Chen, ``{Design and Analysis of Digital Watermarking, Information Embedding,
  and Data Hiding Systems},'' Ph.D. dissertation, Massachusetts Institute of
  Technology, Cambridge, MA, 2000.

\bibitem{chen01:it}
B.~Chen and G.~W. Wornell, ``{Quantization Index Modulation: A Class of
  Provably Good Methods for Digital Watermarking and Information Embedding},''
  \emph{\textit{IEEE Trans. Inform. Theory}}, vol.~47, no.~4, pp. 1423--1443,
  May 2001.

\bibitem{moulin03:it}
P.~Moulin and J.~O'Sullivan, ``{Information-theoretic Analysis of Information
  Hiding},'' \emph{\textit{IEEE Trans. Inform. Theory}}, vol.~49, pp. 563--593,
  2003.

\bibitem{acohen02:it}
A.~S. Cohen, ``{The Gaussian Watermarking Game},'' \emph{\textit{IEEE Trans.
  Inform. Theory}}, vol. vol.48, pp. 1639--1669, June 2002.

\bibitem{caire03:it}
G.~Caire and S.~S. (Shitz), ``{On achievable throughput of a multi-antenna
  Gaussian broadcast channel},'' \emph{\textit{IEEE Trans. Inform. Theory}},
  vol.~49, no.~7, pp. 1691--1706, July 2003.

\bibitem{kalker02:picdsp}
T.~Kalker and F.~Willems, ``{Capacity Bounds and Constructions for Reversible
  Data-hiding},'' in \emph{\textit{Proc. Int. Conf. Digital Signal
  Processing}}, 2002, pp. 71--76.

\bibitem{mitola00:phd}
{J.~Mitola}, ``{Cognitive Radio: an Integrated Agent Architecture for Software
  Defined Radio },'' Ph.D. dissertation, Royal Institute Of Technology,
  Stockholm, Sweden, 2000.

\bibitem{devroye06:it}
N.~Devroye, P.~Mitran, and V.~Tarokh, ``{Achievable Rates in Cognitive
  Channels},'' \emph{\textit{IEEE Trans. Inform. Theory}}, vol.~52, pp.
  1813--1827, May 2006.

\bibitem{jovicic06:it}
A.~Jovicic and P.~Viswanath, ``{Cognitive Radio: An Information-Theoretic
  Perspective},'' submitted to \textit{IEEE Trans. on Information Theory}.

\bibitem{shannon58:ibmj}
C.~E. Shannon, ``{Channels with Side Information at the transmitter},''
  \emph{\textit{IBM J. Res. Devel.}}, vol. vol.2, pp. 289--293, 1958.

\bibitem{salehi92:piee}
M.~Salehi, ``Capacity and coding for memories with real-time noisy defect
  information at encoder and decoder,'' \emph{\textit{Proc. Inst. Elec.
  Eng.-Pt.I.}}, vol. vol.139, pp. 113--117, April 1992.

\bibitem{caire99:it}
G.~Caire and S.~Shamai, ``{On the Capacity of Some Channels with Channel State
  Information},'' \emph{\textit{IEEE Trans. Inform. Theory}}, vol.~45,
  September 1999.

\bibitem{kusnetsov74:ppi}
A.~V. Kusnetsov and B.~S. Tsybakov, ``{Coding in a Memory with Defective
  Cells},'' \emph{\textit{Probl. Peredach. Inform.}}, vol. vol.10, no.~2, pp.
  52--60, April/June 1974.

\bibitem{cheegard83:it}
C.~Heegard and A.~E. Gamal, ``{On the Capacities of Computer Memories with
  Defects},'' \emph{\textit{IEEE Trans. Inform. Theory}}, vol. vol.IT-29, pp.
  731--739, September 1983.

\bibitem{gelfand80:pcit}
S.~I. Gel'fand and M.~S. Pinsker, ``{Coding for Channel with Random
  Parameters},'' \emph{\textit{Probl.Contr. and Information Theory}}, vol.~9,
  no.~1, pp. pp.19--31, 1980.

\bibitem{cover02:it}
T.~Cover and M.~Chiang, ``{Duality Between Channel Capacity and Rate Distortion
  with Two-sided State Information},'' \emph{\textit{IEEE Trans. Inform.
  Theory}}, vol.~48, pp. 1629--1638, June 2002.

\bibitem{mcosta83:it}
M.~H.~M. Costa, ``{Writing on Dirty Paper},'' \emph{\textit{IEEE Trans. Inform.
  Theory}}, vol. vol.IT-29, pp. 439--441, May 1983.

\bibitem{gelfand83:pisit}
S.~I. Gel'fand and M.~S. Pinsker, ``{On Gaussian Channels with Random
  Parameters},'' in \emph{\textit{Proc. {IEEE} Int. Symp. Information Theory
  (ISIT)}}, 1983.

\bibitem{kim04:pisit}
Y.~H. Kim, A.~Sutivong, and S.~Sigurj\'{o}nsson, ``{Multiple User Writing on
  Dirty Paper},'' in \emph{\textit{Proc. {IEEE} Int. Symp. Information Theory
  (ISIT)}}, June 27 - July 2 2004.

\bibitem{baruch06:pallerton}
A.~Somekh-Baruch, S.~Shamai, and S.~Verdu, ``{Cooperative Encoding with
  Asymmetric State Information at the Transmitters},'' in \emph{\textit{Proc.
  Allerton Conf. Communications, Control, and Computing}}, 2006.

\bibitem{baruch07:pisit}
------, ``{Cooperative Multiple Access Encoding with States Available at One
  Transmitter},'' in \emph{\textit{Proc. {IEEE} Int. Symp. Information Theory
  (ISIT)}}, June 24 - June 29 2007.

\bibitem{kotagiri07:pisit}
S.~Kotagiri and J.~N. Laneman, ``{Multiaccess Channels with State Known to One
  Encoder: A Case of Degraded Message Sets},'' in \emph{\textit{Proc. {IEEE}
  Int. Symp. Information Theory (ISIT)}}, June 24 - June 29 2007.

\bibitem{baruch07:it}
A.~Somekh-Baruch, S.~S. (Shitz), and S.~Verdu, ``{Cooperative Multiple Access
  Encoding with States Available at One Transmitter},'' submitted to
  \textit{Tran. Info. Theory}, 2007.

\bibitem{philosof07:pisit}
T.~Philosof, A.~Khisti, U.~Erez, and R.~Zamir, ``{Lattice Strategies for the
  Dirty Multiple Access Channel},'' in \emph{\textit{Proc. {IEEE} Int. Symp.
  Information Theory (ISIT)}}, June 24 - June 29 2007.

\bibitem{ycemal05:it}
Y.~Cemal and Y.~Steinberg, ``{Multiple Access Channel with Partial State
  Information at the Encoders},'' \emph{\textit{IEEE Trans. Inform. Theory}},
  vol. vol.IT-51, pp. 3992--4003, November 2005.

\bibitem{steinberg05:pisit}
Y.~Steinberg and S.~Shamai, ``{Achievable Rates for the Broadcast Channel with
  States Known at the Transmitter},'' in \emph{\textit{Proc. {IEEE} Int. Symp.
  Information Theory (ISIT)}}, September 4 -- September 9 2005.

\bibitem{ysteinberg05:it}
Y.~Steinberg, ``{Coding for the Degraded Broadcast Channel with Random
  parameters, with Causal and Noncausal Side Information},'' \emph{\textit{IEEE
  Trans. Inform. Theory}}, vol. vol.51, pp. 2867--2877, August 2005.

\bibitem{khisti04:pisit}
A.~Khisti, U.~Erez, and G.~W. Wornell, ``{Writing on Many Pieces of Dirty Paper
  at Once: The Binary Case},'' in \emph{\textit{Proc. {IEEE} Int. Symp.
  Information Theory (ISIT)}}, June 27 -- July 2 2004.

\bibitem{kotagiri08:phd}
S.~P. Kotagiri, ``{State-Dependent Networks with Side Information and Partial
  State Recovery },'' Ph.D. dissertation, University of Notre Dame, Notre Dame,
  IN, Dec. 2007.

\bibitem{cover91:book}
T.~M. Cover and J.~A. Thomas, \emph{{Elements of Information Theory}}.\hskip
  1em plus 0.5em minus 0.4em\relax New York: John Wiley \& Sons, Inc., 1991.

\bibitem{zamir02:it}
R.~Zamir, S.~Shamai, and U.~Erez, ``{Nested Linear/Lattice Codes for Structured
  Multiterminal Binning},'' \emph{\textit{IEEE Trans. Inform. Theory}}, vol.
  vol.IT-48, pp. 1250--1276, June 2002.

\bibitem{gallager68:book}
R.~G. Gallager, \emph{{Information ThoTheoryd Reliable Communication}}.\hskip
  1em plus 0.5em minus 0.4em\relax New York: Wiley, 1968.

\bibitem{csiszar81:book}
I.~Csisz\'{a}r and J.~Korner, Eds., \emph{{Information Theory: Coding
  ThoTheoremsr Discrete Memoryless Systems}}.\hskip 1em plus 0.5em minus
  0.4em\relax New York: Academic Press Inc., 1981.

\end{thebibliography}

\end{document}